\documentclass[reprint,aps,prd,onecolumn,notitlepage,showpacs,nofootinbib,preprintnumbers,superscriptaddress]{revtex4-1}
\usepackage{amsmath}
\usepackage{amsfonts}
\usepackage{amssymb}
\usepackage{latexsym}
\usepackage{color}
\usepackage{graphicx}
\usepackage{bm}
\usepackage{epsfig}
\usepackage[english]{babel}
\usepackage{hyperref}
\usepackage{times}
\usepackage{comment}

\begin{document}

\title{Non-canonical two-field inflation to order $\xi^2$}
\author{Yun-Chao Wang}
\author{Towe Wang}
\email[Electronic address: ]{twang@phy.ecnu.edu.cn}
\affiliation{Department of Physics, East China Normal University,\\
Shanghai 200241, China\\ \vspace{0.2cm}}
\date{\today\\ \vspace{1cm}}
\begin{abstract}
In non-canonical two-field inflation models, deviations from the canonical model can be captured by a parameter $\xi$. We show this parameter is usually one half of the slow-roll order and analytically calculate the primordial power spectra accurate to order $\xi^2$. The super-horizon perturbations are studied with an improved method, which gives a correction of order $\xi^2$. Three typical examples demonstrate that our analytical formulae of power spectra fit well with numerical simulation.
\end{abstract}


\maketitle




\section{Motivation}\label{sect-mot}
Popularly taken as one ingredient of the Big Bang theory, inflation \cite{Guth:1980zm} has become the prevailing paradigm for studying the very early phase of our Universe. However, we are still far from the standard model of inflation. Observationally, inflation will be further tested or revealed in details by accumulating data of the cosmic microwave background, the baryon acoustic oscillation, etc \cite{Huang:2015cke}. Theoretically, a consensus has not been attained on the inflationary Lagrangian. Besides the simplest single-field model of inflation \cite{Linde:1981mu,Albrecht:1982wi} and its variants \cite{Starobinsky:1980te,Kehagias:2013mya}, there are a large number of extended models, among which we will study in this paper the generalized two-field inflation model \cite{Polarski:1992dq,GarciaBellido:1995qq,Langlois:1999dw,Starobinsky:2001xq,DiMarco:2005nq,Choi:2007su,Lalak:2007vi,Peterson:2010np,vandeBruck:2014ata} described by an action of the form
\begin{equation}\label{Sin}
S=\int d^{4}x\sqrt{-g}\left[\frac{M_{p}^{2}}{2}R-\frac{1}{2}(\partial_{\mu}\phi)(\partial^{\mu}\phi)-\frac{e^{2b(\phi)}}{2}(\partial_{\mu}\chi)(\partial^{\mu}\chi)-V(\phi,\chi)\right].
\end{equation}
Here $M_{p}=(8\pi G)^{-1/2}$ is the reduced Plank mass, and $b(\phi)$ is a function of scalar field $\phi$. Thus the other scalar field $\chi$ acquires a non-standard kinetic term. In the special case $b=-\phi/(\sqrt{6}M_{p})$, this model is equivalent to the $f(\chi,R)$ generalized gravity \cite{Maeda:1988ab,Nojiri:2003ft,Hwang:2005hb,Ji:2009yw,Saridakis:2016ahq} and the generalized hybrid metric-Palatini gravity \cite{Tamanini:2013ltp}. Another special case $b=\sqrt{2}\phi/(\sqrt{3}M_{p})$ appears in the no-scale supergravity inflation \cite{Ellis:2014gxa,Ellis:2014opa,Chakravarty:2015yho}.

One conventional assumption made in the literature \cite{DiMarco:2005nq,Choi:2007su,Lalak:2007vi,Peterson:2010np} is that $M_{p}^2b_{\phi}^2\ll1$, which unfortunately does not hold in the interesting case $b=-\phi/(\sqrt{6}M_{p})$. For instance, in reference \cite{Lalak:2007vi}, the parameter $\xi=\sqrt{2}b_{\phi}M_{p}\sqrt{\epsilon}$ is introduced and treated ``on the same footing as the other slow-roll parameters'', where $\epsilon$ is the Hubble slow-roll parameter defined in \eqref{sr}. Under this assumption, reference \cite{Lalak:2007vi} kept only the terms linear in $\xi$, $\epsilon$ or $\eta$ and developed a series of elegant formulae for inflation model \eqref{Sin}. Under the same assumption, reference \cite{vandeBruck:2014ata} studied this model to the second order in slow-roll parameters and obtained some analytical results which can be used to find the final power spectra.

Taking $b=-\phi/(\sqrt{6}M_{p})$, one can directly see that $\xi=-\sqrt{\epsilon/3}$. In this case, $\xi$ is of one half order of the slow-roll parameter $\epsilon$, hence $\xi^2$ should be taken on the same footing as $\epsilon$ in a consistent analysis. Our motivation in this paper is to take $\xi^2$ terms into consideration and improve precision of the results of reference \cite{Lalak:2007vi}. We make some refinements to reference \cite{vandeBruck:2014ata}, including the diagonalization of an asymmetric coefficient matrix, the nonzero imaginary part of the correlation spectrum, and an improved treatment of super-horizon perturbations, on which we will comment in section \ref{sect-com}. We make a further step in section \eqref{sect-case} to a subclass of model and work out the power spectra analytically.

The plan of the paper is as follows. In section \ref{sect-back} we write down the background equations as well as the slow-roll conditions. The evolution equation of inflation trajectory in field space is derived with the desired precision in subsection \ref{subsect-traj} and appendix \ref{app-back}. In section \ref{sect-ana}, we evolve the perturbations analytically inside and outside the horizon, taking $\xi^2$ terms into account. The super-horizon power spectra involve integrals difficult to work out analytically, so we pay attention to a subclass of model and perform the analytical calculation in section \ref{sect-case}. For several specific examples of this subclass, our analytical formulae are confronted with numerical results in section \ref{sect-num}. In section \ref{sect-com}, we comment on a few subtleties. Some technical details are relegated to appendices \ref{app-diag} and \ref{app-Hankel}, in which respectively the matrix \eqref{Mtilde} is diagonalized and the Hankel function \eqref{Hexpand} is expanded in the Taylor's series. In appendix \ref{app-int}, we numerically evaluate the integrals involved in the power spectra for specific examples and compare our result with that of reference \cite{vandeBruck:2014ata}. To give some estimate about the contribution of the terms of higher than $\xi^2$ order, we report some analytical results of the order $\xi^{3}$, $\epsilon\xi$, $\eta\xi$ in appendix \ref{app-o3}.

Throughout this paper, we will be restricted to the case with a constant $b_{\phi}$, namely $b_{\phi\phi}=0$. This is enough to cover many popular models, e.g. the $f(\chi,R)$ generalized gravity \cite{Maeda:1988ab,Nojiri:2003ft,Hwang:2005hb,Ji:2009yw,Saridakis:2016ahq}, the generalized hybrid metric-Palatini gravity \cite{Tamanini:2013ltp} and the no-scale supergravity inflation \cite{Ellis:2014gxa,Ellis:2014opa}.

\section{Background equations and slow-roll conditions}\label{sect-back}
In this section we will investigate the evolution of the background under the slow-roll approximation. Starting with action \eqref{Sin}, in subsection \ref{subsect-eom} we will write down the background equations and decompose them in the kinematic basis \cite{Gao:2013zga}. In subsection \ref{subsect-sr}, we will scrutinize the slow-roll condition and put forward a problem related to $f(\chi,R)$ inflation. In subsection \ref{subsect-traj} and appendix \ref{app-back}, we will derive the evolution equations for slow-roll parameters and inflation trajectory after taking $\xi^2$ terms into account.

\subsection{Background equations of motion}\label{subsect-eom}
With the background metric
\begin{eqnarray}
&&ds^{2}=-dt^{2}+a(t)^{2}d\vec{x}^{2},
\end{eqnarray}
the action \eqref{Sin} leads to the following equations of motion for the scalar fields
\begin{eqnarray}
&&\ddot{\phi}+3H\dot{\phi}+V_{\phi}=b_{\phi}e^{2b}\dot{\chi}^{2},\label{phieom}\\
&&\ddot{\chi}+(3H+2b_{\phi}\dot{\phi})\dot{\chi}+e^{-2b}V_{\chi}=0\label{chieom}
\end{eqnarray}
and the Friedmann equations
\begin{eqnarray}
&&H^{2}=\frac{1}{3M_{p}^{2}}\left[\frac{1}{2}\dot{\phi}^{2}+\frac{e^{2b}}{2}\dot{\chi}^{2}+V\right],\label{Fried1}\\
&&\dot{H}=-\frac{1}{2M_{p}^{2}}\left[\dot{\phi}^{2}+e^{2b}\dot{\chi}^{2}\right].\label{Fried2}
\end{eqnarray}
The last equation is not independent from the others. Here $t$ denotes the physical time, and a dot denotes the derivative with respect to it. Later on, we will need the conformal time $\tau$ with respect to which the derivatives are denoted by primes.

To study the evolution of perturbations, it is more convenient to introduce the average and orthogonal fields, corresponding to tangent and orthogonal directions of the trajectory in field space. Perturbations of the average and orthogonal fields $\delta\sigma$, $\delta s$ are related to perturbations of the primitive fields $\delta\phi$, $\delta\chi$ by
\begin{equation}\label{dsigds}
\delta\sigma\equiv\cos\theta\delta\phi+\sin\theta e^{b}\delta\chi,~~~~\delta s\equiv-\sin\theta\delta\phi+\cos\theta e^{b}\delta\chi,
\end{equation}
where
\begin{equation}
\cos\theta\equiv\frac{\dot{\phi}}{\dot{\sigma}},~~~~\quad \sin\theta\equiv\frac{\dot{\chi}e^{b}}{\dot\sigma},~~~~\dot\sigma\equiv\sqrt{\dot{\phi}^{2}+e^{2b}\dot{\chi}^{2}}.
\end{equation}
In terms of average field $\sigma$ and rotation angle $\theta$, equations \eqref{phieom} and \eqref{chieom} can be put in the form \cite{Lalak:2007vi}
\begin{eqnarray}
&&\ddot{\sigma}+3H\dot{\sigma}+V_{\sigma}=0,\label{sigeom}\\
&&\dot{\theta}=-\frac{V_{s}}{\dot{\sigma}}-b_{\phi}\dot{\sigma}\sin\theta.\label{theteom}
\end{eqnarray}

By definition \eqref{dsigds}, the first-order derivatives of the potential are related by
\begin{eqnarray}\label{dpV}
\nonumber&&V_{\sigma}=V_{\phi}c_{\theta}+e^{-b}V_{\chi}s_{\theta},\\
&&V_{s}=-V_{\phi}s_{\theta}+e^{-b}V_{\chi}c_{\theta},
\end{eqnarray}
and the second-order derivatives of the potential are related by
\begin{eqnarray}\label{2dpV}
\nonumber&&V_{\sigma\sigma}=V_{\phi\phi}c_{\theta}^{2}+e^{-2b}V_{\chi\chi}s_{\theta}^{2}+e^{-b}V_{\phi\chi}s_{\theta}c_{\theta}+e^{-b}V_{\chi\phi}s_{\theta}c_{\theta},\\
\nonumber&&V_{\sigma s}=-V_{\phi\phi}s_{\theta}c_{\theta}+e^{-2b}V_{\chi\chi}s_{\theta}c_{\theta}+e^{-b}V_{\phi\chi}c_{\theta}^{2}-e^{-b}V_{\chi\phi}s_{\theta}^{2},\\
&&V_{ss}=V_{\phi\phi}s_{\theta}^{2}+e^{-2b}V_{\chi\chi}c_{\theta}^{2}-e^{-b}V_{\phi\chi}s_{\theta}c_{\theta}-e^{-b}V_{\chi\phi}s_{\theta}c_{\theta}.
\end{eqnarray}
Hereafter we will use notations $s_{\theta}\equiv\sin\theta$, $c_{\theta}\equiv\cos\theta$ for brevity.

One should be warned that the average and orthogonal fields are defined perturbatively in equation \eqref{dsigds}, therefore the potential $V$ cannot be expressed as functions of these fields, and consequently $V_{\sigma}\neq dV/d\sigma$, $V_{\sigma\sigma}\neq dV_{\sigma}/d\sigma$, etc. Instead, a careful calculation shows that
\begin{eqnarray}
\nonumber\dot{V_{\sigma}}&=&V_{\sigma\sigma}\dot{\sigma}+V_{s}\dot{\theta}-b_{\phi}\dot{\sigma}s_{\theta}c_{\theta}(V_{\sigma}s_{\theta}+V_{s}c_{\theta}),\\
\dot{V_{s}}&=&V_{\sigma s}\dot{\sigma}-V_{\sigma}\dot{\theta}-b_{\phi}\dot{\sigma}c_{\theta}^{2}(V_{\sigma}s_{\theta}+V_{s}c_{\theta}),
\end{eqnarray}
where we have made use of $V_{\phi\chi}=V_{\chi\phi}$. Then the time-derivative of equations \eqref{sigeom} and \eqref{theteom} leads to
\begin{eqnarray}\label{3eom}
\nonumber\frac{\dddot{\sigma}}{H^2\dot{\sigma}}&=&3\left(3+\frac{V_{\sigma}}{H\dot{\sigma}}\right)-\frac{3\dot{H}}{H^2}-\frac{V_{\sigma\sigma}}{H^2}+\frac{V_{s}}{H\dot{\sigma}}\left(\frac{V_{s}}{H\dot{\sigma}}+\frac{b_{\phi}\dot{\sigma}}{H}s_{\theta}\right)+\frac{b_{\phi}\dot{\sigma}}{H}s_{\theta}c_{\theta}\left(\frac{V_{\sigma}}{H\dot{\sigma}}s_{\theta}+\frac{V_{s}}{H\dot{\sigma}}c_{\theta}\right),\\
\frac{\ddot{\theta}}{H^2}&=&\left(\frac{b_{\phi}\dot{\sigma}}{H}s_{\theta}-\frac{V_{s}}{H\dot{\sigma}}\right)\left(3+\frac{V_{\sigma}}{H\dot{\sigma}}\right)-\frac{V_{\sigma s}}{H^2}+\left(\frac{b_{\phi}\dot{\sigma}}{H}c_{\theta}-\frac{V_{\sigma}}{H\dot{\sigma}}\right)\left(\frac{V_{s}}{H\dot{\sigma}}+\frac{b_{\phi}\dot{\sigma}}{H}s_{\theta}\right)+\frac{b_{\phi}\dot{\sigma}}{H}c_{\theta}^{2}\left(\frac{V_{\sigma}}{H\dot{\sigma}}s_{\theta}+\frac{V_{s}}{H\dot{\sigma}}c_{\theta}\right).
\end{eqnarray}
This system of equations will be solved in appendix \ref{app-back} order by order in slow-roll parameters defined in the next subsection.

\subsection{Slow-roll conditions and the $\xi$ problem}\label{subsect-sr}
Similar to the single-field inflation, it is customary to define the slow-roll parameters as \cite{DiMarco:2005nq}
\begin{eqnarray}\label{sr}
\nonumber&&\epsilon=-\frac{\dot{H}}{H^{2}},\\
\nonumber&&\eta_{\phi\phi}=\frac{M_{p}^{2}V_{\phi\phi}}{V},~~~~\eta_{\phi\chi}=\frac{M_{p}^{2}e^{-b}V_{\phi\chi}}{V},~~~~\eta_{\chi\chi}=\frac{M_{p}^{2}e^{-2b}V_{\chi\chi}}{V},\\
&&\eta_{\sigma\sigma}=\frac{M_{p}^{2}V_{\sigma\sigma}}{V},~~~~\eta_{\sigma s}=\frac{M_{p}^{2}V_{\sigma s}}{V},~~~~\eta_{ss}=\frac{M_{p}^{2}V_{ss}}{V}.
\end{eqnarray}
Then the slow-roll conditions are
\begin{eqnarray}
&&\epsilon\ll1,\label{srH}\\
&&|\eta_{\phi\phi}|\ll1,~~~~|\eta_{\phi\chi}|\ll1,~~~~|\eta_{\chi\chi}|\ll1,\label{srhi}\\
&&|\eta_{\sigma\sigma}|\ll1,~~~~|\eta_{\sigma s}|\ll1,~~~~|\eta_{ss}|\ll1.\label{srs}
\end{eqnarray}
In terms of slow-roll parameters, relation \eqref{2dpV} can be reexpressed as
\begin{eqnarray}\label{eta}
\nonumber&&\eta_{\sigma\sigma}=\eta_{\phi\phi}c_{\theta}^{2}+\eta_{\chi\chi}s_{\theta}^{2}+2\eta_{\phi\chi}s_{\theta}c_{\theta},\\
\nonumber&&\eta_{\sigma s}=-\eta_{\phi\phi}s_{\theta}c_{\theta}+\eta_{\chi\chi}s_{\theta}c_{\theta}+\eta_{\phi\chi}\left(c_{\theta}^{2}-s_{\theta}^{2}\right),\\
&&\eta_{ss}=\eta_{\phi\phi}s_{\theta}^{2}+\eta_{\chi\chi}c_{\theta}^{2}-2\eta_{\phi\chi}s_{\theta}c_{\theta}.
\end{eqnarray}
Thus the slow-roll conditions \eqref{srhi} and \eqref{srs} are equivalent.

However, as we will see in the next section, the above slow-roll parameters are not enough to express the coefficients of perturbation equations even under the slow-roll approximation. To overcome this difficulty, reference \cite{Lalak:2007vi} introduced a parameter
\begin{equation}\label{xi}
\xi=\frac{b_{\phi}\dot{\sigma}}{H}=\sqrt{2}b_{\phi}M_{p}\sqrt{\epsilon}
\end{equation}
and treated it ``on the same footing as the other slow-roll parameters''. Under this assumption, reference \cite{Lalak:2007vi} kept only terms linear in $\xi$, $\epsilon$ or $\eta$ and developed a series of elegant formulae for inflation model \eqref{Sin}.

By this definition, $\xi$ is proportional to $\sqrt{\epsilon}$, i.e. a half order of slow roll. In theories such the $f(\chi,R)$ generalized gravity \cite{Maeda:1988ab,Nojiri:2003ft,Hwang:2005hb,Ji:2009yw,Saridakis:2016ahq} and the generalized hybrid metric-Palatini gravity \cite{Tamanini:2013ltp}, one has to deal with models of the form $b=-\phi/(\sqrt{6}M_{p})$. In that case $\xi=-\sqrt{\epsilon/3}$, and $\xi^2$ is of order $\epsilon$. In the no-scale supergravity inflation \cite{Ellis:2014gxa,Ellis:2014opa}, $b=\sqrt{2}\phi/(\sqrt{3}M_{p})$. Then one finds $\xi=\sqrt{4\epsilon/3}$ and again $\xi^2$ is of order $\epsilon$. Therefore, it would be interesting to take $\xi^2$ terms into consideration and refine the analysis of reference \cite{Lalak:2007vi,vandeBruck:2014ata}. This will be done in the following sections.

Throughout this paper, we will assume that the slow-roll parameters vary slowly during inflation. This is supported by the differential equations in the coming subsection and is confirmed by our numerical examples in section \ref{sect-num}.

\subsection{Inflation trajectory}\label{subsect-traj}
In section \ref{sect-case}, we will express the power spectra in terms of the Hubble parameter explicitly. For this purpose, we will have to work out the evolution of rotation angle $\theta$ and slow-roll parameters $\epsilon,\eta,\xi$. This has been done to the leading order of $\xi$ in the appendix of reference \cite{vandeBruck:2014ata}. In this subsection, we will extend the results of reference \cite{vandeBruck:2014ata} to $\mathcal{O}\left(\xi^{2}\right)$, assuming $b_{\phi\phi}=0$ and that $\xi^2$ and $\epsilon$ are of the same order.

The evolution of rotation angle $\theta$ is dictated by equation \eqref{theta} in appendix \ref{app-back}. Accurate to $\mathcal{O}\left(\xi^{2}\right)$, it can be rewritten as
\begin{equation}\label{dthet}
\dot{\theta}=-H\left(\eta_{\sigma s}+\xi s_{\theta}c_{\theta}^{2}-\frac{2}{3}\xi^{2}s_{\theta}^{3}c_{\theta}\right).
\end{equation}
Using this equation, one can straightforwardly derive the evolution equation of slow-roll parameters $\epsilon,\eta,\xi$ from their definitions. Since the calculation follows reference \cite{vandeBruck:2014ata} closely, here we write down our final result directly
\begin{eqnarray}\label{deta}
\nonumber\dot{\epsilon}&=&2H\epsilon\left(\epsilon+\frac{\ddot{\sigma}}{H\dot{\sigma}}\right)\simeq 2H\epsilon\left(2\epsilon-\eta_{\sigma\sigma}-\xi s_{\theta}^{2}c_{\theta}-\frac{2}{3}\xi^{2}s_{\theta}^{2}c_{\theta}^{2}\right),\\
\nonumber\dot{\xi}&=&H\xi\left(\epsilon+\frac{\ddot{\sigma}}{H\dot{\sigma}}\right)\simeq H\xi\left(2\epsilon-\eta_{\sigma\sigma}-\xi s_{\theta}^{2}c_{\theta}-\frac{2}{3}\xi^{2}s_{\theta}^{2}c_{\theta}^{2}\right),\\
\nonumber\dot{\eta_{\sigma\sigma}}&=&2H\epsilon\eta_{\sigma\sigma}-2H\eta_{\sigma s}^{2}-2H\eta_{\sigma\sigma}\xi s_{\theta}^{2}c_{\theta}-4H\eta_{\sigma s}\xi s_{\theta}c_{\theta}^{2}+\frac{4}{3}\eta_{\sigma s}\xi^{2}s_{\theta}^{3}c_{\theta}-H\alpha_{\sigma\sigma\sigma},\\
\nonumber\dot{\eta_{ss}}&=&2H\epsilon\eta_{ss}+2H\eta_{\sigma s}^{2}-2H\eta_{ss}\xi c_{\theta}^{3}-\frac{4}{3}\eta_{\sigma s}\xi^{2}s_{\theta}^{3}c_{\theta}-H\alpha_{\sigma ss},\\
\dot{\eta_{\sigma s}}&=&2H\epsilon\eta_{\sigma s}+H\eta_{\sigma s}\eta_{\sigma\sigma}-H\eta_{\sigma s}\eta_{ss} -2H\eta_{ss}\xi s_{\theta}c_{\theta}^{2}-H\eta_{\sigma s}\xi c_{\theta}-\frac{2}{3}\left(\eta_{\sigma\sigma}-\eta_{ss}\right)\xi^{2}s_{\theta}^{3}c_{\theta}-H\alpha_{\sigma\sigma s}.
\end{eqnarray}
In the above $\alpha_{IJK}=V_{\sigma}V_{IJK}/V^2$, whereas the third-order derivatives are \cite{Lalak:2007vi}
\begin{eqnarray}
\nonumber V_{\sigma\sigma\sigma}&=& V_{\phi\phi\phi}c_{\theta}^{3}+3e^{-b}V_{\phi\phi\chi}s_{\theta}c_{\theta}^{2}+3e^{-2b}V_{\phi\chi\chi}s_{\theta}^{2}c_{\theta}+e^{-3b}V_{\chi\chi\chi}s_{\theta}^{3},\\
\nonumber V_{\sigma\sigma s}&=& -V_{\phi\phi\phi}s_{\theta}c_{\theta}^{2}+e^{-b}V_{\phi\phi\chi}c_{\theta}^{3}-2e^{-b}V_{\phi\phi\chi}s_{\theta}^{2}c_{\theta}+2e^{-2b}V_{\phi\chi\chi}s_{\theta}c_{\theta}^{2}-e^{-2b}V_{\phi\chi\chi}s_{\theta}^{3}+e^{-3b}V_{\chi\chi\chi}s_{\theta}^{2}c_{\theta},\\
V_{\sigma ss}&=&V_{\phi\phi\phi}s_{\theta}^{2}c_{\theta}-2e^{-b}V_{\phi\phi\chi}s_{\theta}c_{\theta}^{2}+e^{-b}V_{\phi\phi\chi}s_{\theta}^{3}+e^{-2b}V_{\phi\chi\chi}c_{\theta}^{3}-2e^{-2b}V_{\phi\chi\chi}s_{\theta}^{2}c_{\theta}+e^{-3b}V_{\chi\chi\chi}s_{\theta}c_{\theta}^{2}.
\end{eqnarray}

\section{Analytical study of perturbations including $\xi^2$ terms}\label{sect-ana}
\subsection{Perturbation equations}\label{subsect-pert}
In the longitudinal gauge, the metric with the scalar-type perturbation is of the form
\begin{equation}
ds^{2}=-(1+2\Phi)dt^{2}+a^{2}(1-2\Phi)d\vec{x}^{2},
\end{equation}
while the curvature and entropy perturbations are defined by
\begin{equation}\label{RS}
\mathcal{R}\equiv\frac{H}{\dot{\sigma}}Q_{\sigma},~~~~\mathcal{S}\equiv\frac{H}{\dot{\sigma}}\delta s
\end{equation}
in which
\begin{equation}
Q_{\sigma}\equiv\delta\sigma+\frac{\dot{\sigma}}{H}\Phi.
\end{equation}

In reference \cite{Lalak:2007vi}, it has been shown that perturbations $Q_{\sigma}$ and $\delta s$ are subject to the constraint
\begin{equation}\label{pert0}
\dot{\sigma}\dot{Q}_{\sigma}+\left(3H+\frac{\dot{H}}{H}\right)\dot{\sigma}Q_{\sigma}+V_{\sigma}Q_{\sigma}+2V_{s}\delta s=-\frac{2M_{p}^{2}k^{2}}{a^{2}}\Phi
\end{equation}
and the equations of motion
\begin{eqnarray}
&&\ddot{Q}_{\sigma}+3H\dot{Q}_{\sigma}+\left(\frac{k^{2}}{a^{2}}+C_{\sigma\sigma}\right)Q_{\sigma}+\frac{2V_{s}}{\dot{\sigma}}\dot{\delta s}+C_{\sigma s}\delta s=0,\label{pert1}\\
&&\ddot{\delta s}+3H\dot{\delta s}+\left(\frac{k^{2}}{a^{2}}+C_{ss}\right)\delta s-\frac{2V_{s}}{\dot{\sigma}}\dot{Q}_{\sigma}+C_{s\sigma}Q_{\sigma}=0,\label{pert2}
\end{eqnarray}
where the coefficients are
\begin{eqnarray}\label{C}
\nonumber&&C_{\sigma\sigma}=V_{\sigma\sigma}-\left(\frac{V_{s}}{\dot{\sigma}}\right)^{2}+\frac{2\dot{\sigma} V_{\sigma}}{M_{p}^{2}H}+\frac{3\dot{\sigma}^{2}}{M_{p}^{2}}-\frac{\dot{\sigma}^{4}}{2M_{p}^{4}H^{2}}-b_{\phi}\left[s_{\theta}^{2}c_{\theta}V_{\sigma}+(c_{\theta}^{2}+1)s_{\theta}V_{s}\right],\\
\nonumber&&C_{\sigma s}=6H\frac{V_{s}}{\dot{\sigma}}+\frac{2V_{\sigma}V_{s}}{\dot{\sigma}^{2}}+2V_{\sigma s}+\frac{\dot{\sigma}V_{s}}{M_{p}^{2}H}+2b_{\phi}\left(s_{\theta}^{3}V_{\sigma}-c_{\theta}^{3}V_{s}\right),\\
\nonumber&&C_{ss}=V_{ss}-\left(\frac{V_{s}}{\dot{\sigma}}\right)^{2}+b_{\phi}(1+s_{\theta}^{2})c_{\theta}V_{\sigma}+b_{\phi}s_{\theta}c_{\theta}^{2}V_{s}-b_{\phi}^{2}\dot{\sigma}^{2},\\
&&C_{s\sigma}=-6H\frac{V_{s}}{\dot{\sigma}}-\frac{2V_{\sigma}V_{s}}{\dot{\sigma}^{2}}+\frac{\dot{\sigma} V_{s}}{M_{p}^{2}H}.
\end{eqnarray}
In terms of slow-roll parameters $\epsilon$, $\eta$ and $\xi$, these coefficients can be rewritten under the slow-roll approximation as
\begin{eqnarray}\label{Csr}
\nonumber C_{\sigma\sigma}&=&H^{2}\left(3\xi s_{\theta}^{2}c_{\theta}-6\epsilon+3\eta_{\sigma\sigma}+\xi^{2}s_{\theta}^{4}c_{\theta}^{2}\right)+\mathcal{O}\left(\epsilon^{2},\eta^{2},\xi^{3},\epsilon\eta,\epsilon\xi,\eta\xi\right),\\
\nonumber C_{\sigma s}&=&H^{2}(-6\xi s_{\theta}^{3}+6\eta_{\sigma s}+2\xi^{2}s_{\theta}^{3}c_{\theta}^{3})+\mathcal{O}\left(\epsilon^{2},\eta^{2},\xi^{3},\epsilon\eta,\epsilon\xi,\eta\xi\right),\\
\nonumber C_{ss}&=&H^{2}\left[-3\xi c_{\theta}(1+s_{\theta}^{2})+3\eta_{ss}+\xi^{2}\left(s_{\theta}^{4}c_{\theta}^{2}-c_{\theta}^{2}-2s_{\theta}^{4}\right)\right]+\mathcal{O}\left(\epsilon^{2},\eta^{2},\xi^{3},\epsilon\eta,\epsilon\xi,\eta\xi\right),\\
C_{s\sigma}&=&2\xi^{2}s_{\theta}^{5}c_{\theta}+\mathcal{O}\left(\epsilon^{2},\eta^{2},\xi^{3},\epsilon\eta,\epsilon\xi,\eta\xi\right).
\end{eqnarray}
We mention in passing that equations \eqref{C} and \eqref{Csr} match equations (2.16-2.19) and (2.29-2.32) in reference \cite{vandeBruck:2014ata} except for the third equation of \eqref{Csr}. The third term in the square brackets of this equation is replaced by $\xi^{2}c_{\theta}^{2}\left(s_{\theta}^{4}-1\right)$ in reference \cite{vandeBruck:2014ata}.  Such a small difference will give rise to a difference in the power spectra, as will be explained in subsection \ref{subsect-sup} and illustrated in section \ref{sect-num}.

With the perturbation equations of motion \eqref{pert1} and \eqref{pert2} at hand, we will study the evolution of perturbations $Q_{\sigma}$, $\delta s$ along the timeline, where the physical wavelength $a/k$ grows together with the scale factor $a$. At the very beginning, the physical wavenumber is much bigger than the Hubble parameter $k/a\gg H$. Therein  $aQ_{\sigma}$, $a\delta s$ behave as free-field fluctuations, living in a Minkowski vacuum. We will discuss the Minkowski-like vacuum as initial conditions in subsection \ref{subsect-init}. Starting from such initial conditions, the evolution of perturbations during inflation can be roughly divided into two stages: the sub-horizon stage $k/(aH)\gtrsim1$ and the super-horizon stage $k/(aH)\lesssim1$. Perturbation evolution during the two stages will be investigated in subsections \ref{subsect-sub} and \ref{subsect-sup}. It is customary to mark the timeline with the number of e-folds. In our convention of notations, the number of e-folds after Hubble crossing is defined by
\begin{equation}
N=\ln\frac{a}{a_{\ast}}.
\end{equation}
Hereafter the quantities with a subscript star are evaluated at Hubble crossing $k=a_{\ast}H_{\ast}$.

\subsection{Initial conditions}\label{subsect-init}
Before studying the evolution of perturbations, we should set their initial conditions. At the very beginning of inflation, the physical wavelength $a/k$ is far smaller than the radius of Hubble horizon $1/H$, i.e. $k/a\gg H$, so the coupling terms and mass terms become negligible in equations \eqref{pert1} and \eqref{pert2}, which reduce to the equations of motion of free harmonic oscillators. Then as initial conditions, the fluctuations $aQ_{\sigma}$ and $a\delta s$ can be quantized as free fields. That is to say, at the initial time $\tau_{i}\rightarrow-\infty$, we have \cite{Gordon:2000hv}
\begin{equation}\label{init}
Q_{\sigma}(\tau_{i})\simeq\frac{ e^{-\mathrm{i}k\tau_{i}}}{a(\tau_{i})\sqrt{2k}}e_{\sigma}(k),~~~~\delta s(\tau_{i})\simeq\frac{e^{-\mathrm{i}k\tau_{i}}}{a(\tau_{i})\sqrt{2k}}e_{s}(k).
\end{equation}
Here the independent Gaussian random variables $e_{\sigma}$ and $e_{s}$ are orthogonally normalized
\begin{equation}
\langle e_{I}(k)\rangle=0,~~~~\langle e_{I}(k)e^{\ast}_{J}(k')\rangle=\delta_{IJ}\delta^{(3)}(k-k')
\end{equation}
with $I,J=\sigma,s$.

In the next subsection, at Hubble crossing, we will need two other Gaussian random variables $e_{1}$ and $e_{2}$, which are related to $e_{\sigma}$ and $e_{s}$ via
\begin{equation}
\left(\begin{array}{c}
e_{1} \\
e_{2}\end{array}\right)=\left(\begin{array}{cc}
                              \cos\Theta_{\ast} & \sin\Theta_{\ast} \\
                              \sin\Psi_{\ast} & \cos\Psi_{\ast}
                            \end{array}\right)\left(\begin{array}{c}
                                                    e_{\sigma} \\
                                                    e_{s}
                                                  \end{array}\right),
\end{equation}
Note that $e_{1}$, $e_{2}$, $e_{\sigma}$, $e_{s}$ are time-independent variables, and $\Theta_{\ast}$, $\Psi_{\ast}$ are also time-independent. This transformation matrix generalizes the rotation matrix in references \cite{Lalak:2007vi,Byrnes:2006fr}. Especially, when $\Psi_{\ast}=-\Theta_{\ast}$, it reduces to the relation in \cite{Lalak:2007vi}. In our general form, $e_{1}$ and $e_{2}$ are not always orthogonal to each other. Instead,
\begin{equation}\label{nonorth}
\langle e_{A}(k)e^{\ast}_{B}(k')\rangle=\Delta_{AB}\delta^{(3)}(k-k')
\end{equation}
with $A,B=1,2$ and
\begin{equation}\label{Del}
\Delta_{AB}\equiv\left\{\begin{array}{ll}
                1, & A=B; \\
                \sin(\Theta_{\ast}+\Psi_{\ast}), & A \neq B.
              \end{array}\right.
\end{equation}

\subsection{Sub-horizon evolution and horizon crossing}\label{subsect-sub}
In this subsection, we will study the dynamical evolution of perturbations $Q_{\sigma}$ and $\delta s$ from deep inside the Hubble horizon to Hubble crossing. Introducing variables
\begin{equation}
u_{\sigma}=aQ_{\sigma},~~~~u_{s}=a\delta s,
\end{equation}
and using the conformal time $\tau$, we can rewrite equations \eqref{pert1} and \eqref{pert2} as
\begin{eqnarray}
\nonumber&&u_{\sigma}''+\frac{2V_{s}}{\dot{\sigma}}a u_{s}'+\left[k^{2}-\frac{a''}{a}+a^{2}C_{\sigma\sigma}\right]u_{\sigma}+\left[-\frac{2V_{s}}{\dot{\sigma}}a'+a^{2}C_{\sigma s}\right]u_{s}=0,\\
&&u_{s}''-\frac{2V_{s}}{\dot{\sigma}}a u_{\sigma}'+\left[k^{2}-\frac{a''}{a}+a^{2}C_{ss}\right]u_{s}+\left[\frac{2V_{s}}{\dot{\sigma}}a'+a^{2}C_{s\sigma}\right]u_{\sigma}=0.
\end{eqnarray}
The four coefficients $C_{IJ}$ have been given by equations \eqref{C}. Unlike reference \cite{Lalak:2007vi}, in the following discussion, we will keep terms of $\mathcal{O}\left(\epsilon,\eta,\xi^{2}\right)$.

From definition $d \tau=dt/a$, we can get some useful relations
\begin{eqnarray}
\nonumber&&\tau=-\frac{1+\epsilon}{Ha},~~~~a'=a^{2}H,\\
&&\frac{1}{a^{2}}\simeq H^2\tau^2(1-2\epsilon),~~~~\frac{a''}{a}\simeq\frac{2+3\epsilon}{\tau^{2}},
\end{eqnarray}
which will be utilized without mention hereafter. Then the perturbation equations can be put in the matrix form
\begin{equation}
\left[\left(\frac{d^{2}}{d\tau^{2}}+k^{2}-\frac{2+3\epsilon}{\tau^{2}}\right)\mathbf{I}+2\mathbf{E}\frac{1}{\tau}\frac{d}{d\tau}+\mathbf{M}\frac{1}{\tau^{2}}\right]\left(\begin{array}{c}
                                                                                                                                                                          u_{\sigma}\\
                                                                                                                                                                          u_{s}
                                                                                                                                                                          \end{array}\right)=0,
\end{equation}
where $\mathbf{I}$ is the unitary matrix, and matrices $\mathbf{E}$, $\mathbf{M}$ are
\begin{eqnarray}\label{EM}
\nonumber&&\mathbf{E}=\left(\begin{array}{cc}
                0 & -(1+\epsilon)\left(\eta_{\sigma s}-\xi s_{\theta}^{3}-\frac{2}{3}\xi^{2}s_{\theta}^{3}c_{\theta}\right) \\
                (1+\epsilon)\left(\eta_{\sigma s}-\xi s_{\theta}^{3}-\frac{2}{3}\xi^{2}s_{\theta}^{3}c_{\theta}\right) & 0 \\
              \end{array}\right),\\
&&\mathbf{M}=\left(\begin{array}{cc}
                 H^{-2}C_{\sigma\sigma}(1+\epsilon)^{2} & \left(-2\eta_{\sigma s}+2\xi s_{\theta}^{3}+\frac{4}{3}\xi^{2}s_{\theta}^{3}c_{\theta}+H^{-2}C_{\sigma s}\right)(1+\epsilon)^{2} \\
                 \left(2\eta_{\sigma s}-2\xi s_{\theta}^{3}-\frac{4}{3}\xi^{2}s_{\theta}^{3}c_{\theta}+H^{-2}C_{s\sigma}\right)(1+\epsilon)^{2} & H^{-2}C_{ss}(1+\epsilon)^{2} \\
               \end{array}\right).
\end{eqnarray}
The above system of perturbation equations can be succinctly written as
\begin{equation}
u''+2\mathbf{L}u'+\mathbf{Q}u=0
\end{equation}
if we take notations
\begin{equation}
u=\left(\begin{array}{c}
    u_{\sigma}\\
    u_{s}
    \end{array}\right),~~~~\mathbf{L}=\frac{1}{\tau}\mathbf{E},~~~~\mathbf{Q}=\left(k^{2}-\frac{2+3\epsilon}{\tau^{2}}\right)\mathbf{I}+\frac{1}{\tau^{2}}\mathbf{M}.
\end{equation}

It is remarkable that matrices $\mathbf{E}$, $\mathbf{L}$ here are exactly the same as them in reference \cite{Lalak:2007vi}. As argued in reference \cite{Lalak:2007vi}, there is always an orthogonal matrix $\mathbf{R}$ obeying the differential equation
\begin{equation}
\frac{d}{d\tau}\mathbf{R}=-\mathbf{LR},
\end{equation}
and $\mathbf{R}$ is slowly varying because $\mathbf{E}$ is linear in slow-roll parameters. Consequently, in terms of $v=\mathbf{R}^{-1}u$, the system of perturbation equations can be further rewritten as
\begin{equation}\label{pertv}
v''+\mathbf{R}^{-1}(-\mathbf{L}^{2}-\mathbf{L}'+\mathbf{Q})\mathbf{R}v=0,
\end{equation}
where $\mathbf{L}'=d\mathbf{L}/d\tau$. From relation $\mathbf{L}=\mathbf{E}/\tau$, it is easy to see
\begin{equation}
-\mathbf{L}^{2}-\mathbf{L}'=-\frac{\mathbf{E}^{2}}{\tau^{2}}+\frac{\mathbf{E}}{\tau^{2}}
\end{equation}
if slow-roll parameters vary sufficiently slowly. Here we have kept the $\mathbf{E}^{2}$ term in order to obtain $\mathcal{O}\left(\xi^{2}\right)$ terms.

To proceed, we have to decouple the system \eqref{pertv} into two independent equations. That is equivalent to diagonalizing the coefficient matrix of $v$ through a similarity transformation. For this purpose, we should diagonalize the following matrix
\begin{eqnarray}\label{Mtilde}
\nonumber\tilde{\mathbf{M}}&=&-\mathbf{E^{\mathrm{2}}+E+M}\\
&=&\left(
     \begin{array}{cc}
       3\xi s_{\theta}^{2}c_{\theta}-6\epsilon+3\eta_{\sigma\sigma} & -3\xi s_{\theta}^{3}+3\eta_{\sigma s} \\
       -3\xi s_{\theta}^{3}+3\eta_{\sigma s} & -3\xi c_{\theta}-3\xi s_{\theta}^{2}c_{\theta}+3\eta_{ss} \\
     \end{array}
   \right)+\left(
                     \begin{array}{cc}
                       \xi^{2}s_{\theta}^{4} & 2\xi^{2}s_{\theta}^{3}c_{\theta}^{3}+2\xi^{2}s_{\theta}^{3}c_{\theta} \\
                      2\xi^{2}s_{\theta}^{5}c_{\theta}-2\xi^{2}s_{\theta}^{3}c_{\theta} & -\xi^{2}+\xi^{2}s_{\theta}^{2}c_{\theta}^{2} \\
                     \end{array}
                   \right),
\end{eqnarray}
where we have kept terms of $\mathcal{O}\left(\epsilon,\eta,\xi^{2}\right)$. The second term, which is proportional to $\xi^{2}$, has been neglected in reference \cite{Lalak:2007vi}. In this paper, as we have explained in subsection \ref{subsect-sr}, $\xi^{2}$ is supposed to be of order $\epsilon$ and thus nonnegligible. This makes our investigation more challenging than \cite{Lalak:2007vi}. In particular, due to the second term, the above matrix is no longer symmetric and hence cannot be diagonalized through an orthogonal similarity transformation. Fortunately, since $\tilde{\mathbf{M}}$ has analytically two distinct eigenvalues $\tilde{\lambda}_{1}$, $\tilde{\lambda}_{2}$ as given in appendix \ref{app-diag}, we can still diagonalize the matrix through a similarity transformation. Remember that if there are exactly $n$ distinct eigenvalues in an $n\times n$ matrix, then this matrix is diagonalizable. However, such a similarity transformation is not always an orthogonal transformation. That is why in subsection \ref{subsect-init} the variables $e_{1}$, $e_{2}$ are not necessarily orthogonal to each other. The details are given below.

Equation \eqref{Mtilde} is to be compared with equations (2.39), (2.40) in reference \cite{vandeBruck:2014ata}. Comparing $D_{Q}$ in reference \cite{vandeBruck:2014ata} with the corresponding component here, we find the difference $2\xi^{2}s_{\theta}^{4}$. This can be attributed to the difference in $C_{ss}$ as noted in subsection \ref{subsect-pert}. The off-diagonal terms in reference \cite{vandeBruck:2014ata} are also different from our result here. Unfortunately, reference \cite{vandeBruck:2014ata} did not reveal much calculative details, though we note that $B_{Q}=C_{Q}$ therein is exactly the average value of off-diagonal components here. Such a small difference leads to an asymmetric $\tilde{\mathbf{M}}$, and for the aforementioned reason, makes our following investigation more challenging.

Since the slow-roll parameters are approximately constant, one can diagonalize matrix $\tilde{\mathbf{M}}$ with a time-independent matrix $\tilde{\mathbf{R}}_{\ast}$ as
\begin{equation}\label{diag}
\tilde{\mathbf{R}}_{\ast}^{-1}\tilde{\mathbf{M}}\tilde{\mathbf{R}}_{\ast}=\mathrm{Diag}(\tilde{\lambda}_{1},\tilde{\lambda}_{2}).
\end{equation}
The values of $\tilde{\lambda}_{1}$, $\tilde{\lambda}_{2}$ are given by equations \eqref{eigen}. Without loss of generality and for later convenience, we parameterize the inverse of $\tilde{\mathbf{R}}_{\ast}$ with two angles $\Theta_{\ast}$, $\Psi_{\ast}$ in the following form
\begin{equation}\label{Rinverse}
\tilde{\mathbf{R}}_{\ast}^{-1}=\left(
                            \begin{array}{cc}
                              \cos\Theta_{\ast} & \sin\Theta_{\ast} \\
                              \sin\Psi_{\ast} & \cos\Psi_{\ast} \\
                            \end{array}
                          \right).
\end{equation}
From $\tilde{\mathbf{R}}_{\ast}\tilde{\mathbf{R}}_{\ast}^{-1}=\mathbf{I}$, we can directly write down
\begin{equation}
\tilde{\mathbf{R}}_{\ast}=\frac{1}{\cos(\Theta_{\ast}+\Psi_{\ast})}\left(
                            \begin{array}{cc}
                              \cos\Psi_{\ast} & -\sin\Theta_{\ast} \\
                              -\sin\Psi_{\ast} & \cos\Theta_{\ast} \\
                            \end{array}
                          \right).
\end{equation}

If we further transform $v$ into
\begin{equation}
w=\tilde{\mathbf{R}}_{\ast}^{-1}\mathbf{R}_{\ast}v,
\end{equation}
then because $\mathbf{R}$ varies slowly, we have approximately $w\simeq\tilde{\mathbf{R}}_{\ast}^{-1}u$, and perturbation equations \eqref{pertv} can be decoupled into two independent equations of the form
\begin{equation}
w_{A}''+[k^{2}-\frac{1}{\tau^{2}}(2+3\lambda_{A})]w_{A}=0\label{pertome}
\end{equation}
with $A=1,2$ and
\begin{equation}
\lambda_{A}=\epsilon-\frac{\tilde{\lambda}_{A}}{3}.
\end{equation}
For matrix \eqref{Mtilde}, in appendix \ref{app-diag} we present the eigenvalues $\tilde{\lambda}_{1}$ and $\tilde{\lambda}_{2}$ as well as some trigonometric functions of $\Psi_{\ast}$ and $\Theta_{\ast}$.

Keep in mind that $\tilde{\mathbf{R}}_{\ast}$ is not necessarily an orthogonal matrix. As a result, the solutions $w_1$, $w_2$ of \eqref{pertome} are not orthogonal to each other. This introduces extra complications to our study. To see this, let us utilize the notation
\begin{equation}
\mu_{A}=\sqrt{\frac{9}{4}+3\lambda_{A}}
\end{equation}
to write down the solutions of \eqref{pertome} as \cite{Byrnes:2006fr}
\begin{equation}\label{wsol}
w_{A}=\frac{\sqrt{\pi}}{2}e^{\mathrm{i}(\mu_{A}+1/2)\pi/2}\sqrt{-\tau}H_{\mu_{A}}^{(1)}(-k\tau) e_{A}(k)
\end{equation}
where $H_{\mu}^{(1)}$ is the Hankel function of the first kind of order $\mu$, and the variables $e_{1}$, $e_{2}$ have been defined in subsection \ref{subsect-init}. Clearly $\langle w^{\dag}_{1}w_{2}\rangle\neq0$ because of equation \eqref{nonorth}. This will have significant implications for the calculation of power spectra.

Considering that $w\simeq\tilde{\mathbf{R}}_{\ast}^{-1}u$ at Hubble crossing, we can prove
\begin{eqnarray}
\nonumber aQ_{\sigma}&=&\frac{\cos\Psi_{\ast}}{\cos(\Theta_{\ast}+\Psi_{\ast})} w_{1}-\frac{\sin\Theta_{\ast}}{\cos(\Theta_{\ast}+\Psi_{\ast})} w_{2}, \\
\nonumber a\delta s&=&-\frac{\sin\Psi_{\ast}}{\cos(\Theta_{\ast}+\Psi_{\ast})} w_{1}+\frac{\cos\Theta_{\ast}}{\cos(\Theta_{\ast}+\Psi_{\ast})} w_{2}
\end{eqnarray}
and immediately write down their correlations
\begin{eqnarray}\label{QQ}
\nonumber a^{2}\langle Q_{\sigma}^{\dag}Q_{\sigma}\rangle&=&\frac{\cos^{2}\Psi_{\ast}}{\cos^{2}(\Theta_{\ast}+\Psi_{\ast})}\langle w^{\dag}_{1}w_{1}\rangle+\frac{\sin^{2}\Theta_{\ast}}{\cos^{2}(\Theta_{\ast}+\Psi_{\ast})}\langle w^{\dag}_{2}w_{2}\rangle-\frac{\cos\Psi_{\ast}\sin\Theta_{\ast}}{\cos^{2}(\Theta_{\ast}+\Psi_{\ast})}\left(\langle w^{\dag}_{1}w_{2}\rangle+\langle w^{\dag}_{2}w_{1}\rangle\right),\\
a^{2}\langle {\delta s}^{\dag}\delta s\rangle&=&\frac{\sin^{2}\Psi_{\ast}}{\cos^{2}(\Theta_{\ast}+\Psi_{\ast})}\langle w^{\dag}_{1}w_{1}\rangle+\frac{\cos^{2}\Theta_{\ast}}{\cos^{2}(\Theta_{\ast}+\Psi_{\ast})}\langle w^{\dag}_{2}w_{2}\rangle-\frac{\sin\Psi_{\ast}\cos\Theta_{\ast}}{\cos^{2}(\Theta_{\ast}+\Psi_{\ast})}\left(\langle w^{\dag}_{1}w_{2}\rangle+\langle w^{\dag}_{2}w_{1}\rangle\right),\\
\nonumber a^{2}\langle Q_{\sigma}^{\dag}\delta s\rangle&=&-\frac{\cos\Psi_{\ast}\sin\Psi_{\ast}}{\cos^{2}(\Theta_{\ast}+\Psi_{\ast})}\langle w^{\dag}_{1}w_{1}\rangle
-\frac{\cos\Theta_{\ast}\sin\Theta_{\ast}}{\cos^{2}(\Theta_{\ast}+\Psi_{\ast})}\langle w^{\dag}_{2}w_{2}\rangle+\frac{\sin\Theta_{\ast}\sin\Psi_{\ast}}{\cos^{2}(\Theta_{\ast}+\Psi_{\ast})}\langle w^{\dag}_{2}w_{1}\rangle+\frac{\cos\Psi_{\ast}\cos\Theta_{\ast}}{\cos^{2}(\Theta_{\ast}+\Psi_{\ast})}\langle w^{\dag}_{1}w_{2}\rangle.
\end{eqnarray}
Remember that $Q_{\sigma}$ and $\delta s$ are related to curvature perturbation $\mathcal{R}$ and entropy perturbation $\mathcal{S}$ respectively by \eqref{RS}. With the above correlations, in principle one can compute the power spectra of $\mathcal{R}$ and $\mathcal{S}$ as well as their correlation spectrum by definitions
\begin{eqnarray}\label{RR}
\nonumber\langle \mathcal{R}_{k}^{\ast}\mathcal{R}_{k'}\rangle&=&\frac{2\pi^{2}}{k^{3}}\mathcal{P_{R}}(k)\delta^{(3)}(k-k'),\\
\nonumber\langle \mathcal{S}_{k}^{\ast}\mathcal{S}_{k'}\rangle&=&\frac{2\pi^{2}}{k^{3}}\mathcal{P_{S}}(k)\delta^{(3)}(k-k'),\\
\langle \mathcal{R}_{k}^{\ast}\mathcal{S}_{k'}\rangle&=&\frac{2\pi^{2}}{k^{3}}\mathcal{C_{RS}}(k)\delta^{(3)}(k-k')
\end{eqnarray}
and further more the relative correlation coefficient
\begin{equation}\label{Crel}
\tilde{\mathcal{C}}=\frac{\mathcal{C_{RS}}}{\sqrt{\mathcal{P_{R}P_{S}}}}.
\end{equation}

But in practice the computation is rather complicated when $\Psi_{\ast}\neq-\Theta_{\ast}$ and $\langle w^{\dag}_{1}w_{2}\rangle\neq0$, because we want to keep $\mathcal{O}\left(\xi^{2}\right)$ corrections. To accomplish the task, we will expand the spectra to $\mathcal{O}\left(\epsilon,\eta,\xi^{2}\right)$ at several e-folds after Hubble crossing, that is $k/(aH)\ll1$ or equivalently $-k\tau\ll1$.

From equations \eqref{eigen}, we can see the leading-order term in $\lambda_{A}$ is linear in $\xi$, so in our series expansion below we will keep $\mathcal{O}\left(\lambda_{A}^{2}\right)$ terms. For instance,
\begin{equation}\label{muA}
\mu_{A}\simeq\frac{3}{2}+\lambda_{A}-\frac{1}{3}\lambda_{A}^{2}.
\end{equation}
In terms of the gamma function $\Gamma(\mu)$ and Bessel functions of the first kind $J_{\mu}(x)$, the Hankel function $H_{\mu}^{(1)}(x)$ can be expressed as
\begin{eqnarray}\label{HJ}
\nonumber&&H_{\mu}^{(1)}(x)=\frac{J_{-\mu}(x)-\mathrm{e}^{-\mathrm{i}\mu\pi}J_{\mu}(x)}{\mathrm{i}\sin(\mu\pi)},\\
&&J_{\mu}(x)=\sum^{\infty}_{m=0}\frac{(-1)^{m}}{m!\Gamma(m+\mu+1)}\left(\frac{x}{2}\right)^{2m+\mu}.
\end{eqnarray}
Apparently, when $\mu>0$ and $x\ll1$, the Hankel function $H_{\mu}^{(1)}(x)$ is divergent as $x^{-\mu}$. If we renormalize $H_{\mu_{A}}^{(1)}(x)$ as
\begin{equation}
\tilde{H}_{\mu_{A}}^{(1)}(x)\equiv x^{\mu_{A}}H_{\mu_{A}}^{(1)}(x)
\end{equation}
and take \eqref{muA} into account, then $\tilde{H}_{\mu_{A}}^{(1)}(x)$ can be expanded to $\mathcal{O}\left(\lambda_{A}^{2}\right)$ as
\begin{equation}\label{Hexpand}
\tilde{H}_{\mu_{A}}^{(1)}(x)\simeq\tilde{H}_{3/2}^{(1)}(x)+\left(\lambda_{A}-\frac{\lambda_{A}^{2}}{3}\right)\left.\frac{d\tilde{H}_{\mu}^{(1)}(x)}{d\mu}\right|_{\mu=3/2}+\frac{\lambda_{A}^2}{2}\left.\frac{\mathrm{d^2}\tilde{H}_{\mu}^{(1)}(x)}{d\mu^2}\right|_{\mu=3/2}.
\end{equation}
This leads to
\begin{eqnarray}
\nonumber&&\frac{\pi}{2}x^{\mu_{A}+\mu_{B}}H_{\mu_{A}}^{(1)\ast}(x)H_{\mu_{B}}(x)\\
\nonumber&\simeq&\frac{\pi}{2}x^{3}\left|H_{3/2}^{(1)}(x)\right|^2\left[1+\left(\lambda_{A}-\frac{\lambda_{A}^{2}}{3}\right)f(x)+\frac{\lambda_{A}^2}{2}g(x)\right]\left[1+\left(\lambda_{B}-\frac{\lambda_{B}^{2}}{3}\right)f(x)+\frac{\lambda_{B}^2}{2}g(x)\right]\\
&\simeq&(1+x^{2})\left[1+\left(\lambda_{A}+\lambda_{B}-\frac{\lambda_{A}^{2}}{3}-\frac{\lambda_{B}^{2}}{3}\right)f(x)+\lambda_{A}\lambda_{B}f^2(x)+\frac{1}{2}\left(\lambda_{A}^2+\lambda_{B}^{2}\right)g(x)\right].
\end{eqnarray}
In the expression we have introduced
\begin{eqnarray}
\nonumber f(x)&=&\frac{1}{\tilde{H}_{3/2}^{(1)}(x)}\frac{d\tilde{H}_{\mu}^{(1)}(x)}{d\mu}\bigg |_{\mu=3/2}\\
\nonumber&\simeq&2-\ln2-\gamma-\frac{2x^2}{2+x^2}\\
\nonumber&\simeq&0.7296-x^2,\\
\nonumber g(x)&=&\frac{1}{\tilde{H}_{3/2}^{(1)}(x)}\frac{\mathrm{d^2}\tilde{H}_{\mu}^{(1)}(x)}{d\mu^2}\bigg |_{\mu=3/2}\\
\nonumber&\simeq&\frac{\pi^2}{2}+(\ln2+\gamma)\left(-4+\ln2+\gamma+\frac{4x^2}{2+x^2}\right)\\
&\simeq&1.4672+2.5407x^{2},
\end{eqnarray}
where $\gamma\simeq0.5772$ is the Euler-Mascheroni constant. The details for evaluating $f(x)$ and $g(x)$ are relegated to appendix \ref{app-Hankel}.

With this result at hand, using the modified normalization condition \eqref{nonorth}, we can work out the correlations of the two solutions \eqref{wsol} to $\mathcal{O}\left(\lambda_{A}^{2}\right)$, yielding
\begin{eqnarray}\label{ww}
\nonumber\langle w^{\dag}_{A}(k)w_{B}(k')\rangle&=&\frac{\pi}{4}(-\tau)e^{\mathrm{i}\pi(\mu_{B}-\mu_{A})/2}H_{\mu_{A}}^{\ast}(-k\tau)H_{\mu_{B}}(-k\tau)\Delta_{AB}\delta^{(3)}(k-k')\\
\nonumber&\simeq&e^{\mathrm{i}\pi(\mu_{B}-\mu_{A})/2}\frac{-\tau(1+k^{2}\tau^{2})}{2(-k\tau)^{\mu_{A}+\mu_{B}}}\biggl[1+\left(\lambda_{A}+\lambda_{B}-\frac{\lambda_{A}^{2}}{3}-\frac{\lambda_{B}^{2}}{3}\right)f(-k\tau)+\lambda_{A}\lambda_{B}f^2(-k\tau)\\
&&+\frac{1}{2}\left(\lambda_{A}^2+\lambda_{B}^{2}\right)g(-k\tau)\biggr]\Delta_{AB}\delta^{(3)}(k-k')
\end{eqnarray}
with $\Delta_{AB}$ defined by \eqref{Del}. Note here
\begin{eqnarray}
\nonumber e^{\mathrm{i}\pi(\mu_{B}-\mu_{A})/2}&\simeq&1+\frac{\mathrm{i}\pi}{2}(\mu_{B}-\mu_{A})-\frac{\pi^2}{8}(\mu_{B}-\mu_{A})^2\\
&\simeq&1+\frac{\mathrm{i}\pi}{2}\left(\lambda_{B}-\lambda_{A}-\frac{\lambda_{B}^{2}}{3}+\frac{\lambda_{A}^{2}}{3}\right)-\frac{\pi^2}{8}(\lambda_{B}-\lambda_{A})^2.
\end{eqnarray}

Inserting \eqref{ww} into equations \eqref{QQ}, by definitions \eqref{RR} we finally obtain the spectra after Hubble crossing
\begin{eqnarray}\label{PRstar}
\nonumber\bar{\mathcal{P}}_{\mathcal{R}}&=&\left(\frac{H^2_{\ast}}{2\pi\dot{\sigma}_{\ast}}\right)^{2}(1+k^{2}\tau^{2})(1-2\epsilon_{\ast})\biggl\{\frac{\cos^{2}\Psi_{\ast}+\sin^{2}\Theta_{\ast}}{\cos^{2}(\Theta_{\ast}+\Psi_{\ast})}
+\frac{(6\lambda_{1}-2\lambda_{1}^2)\cos^{2}\Psi_{\ast}+(6\lambda_{2}-2\lambda_{2}^2)\sin^{2}\Theta_{\ast}}{3\cos^{2}(\Theta_{\ast}+\Psi_{\ast})}f(-k\tau)\\
\nonumber&&+\frac{\lambda_{1}^2\cos^{2}\Psi_{\ast}+\lambda_{2}^2\sin^{2}\Theta_{\ast}}{\cos^{2}(\Theta_{\ast}+\Psi_{\ast})}\left[f^2(-k\tau)+g(-k\tau)\right]-\frac{\cos\Psi_{\ast}\sin\Theta_{\ast}}{\cos^{2}(\Theta_{\ast}+\Psi_{\ast})}\sin(\Theta_{\ast}+\Psi_{\ast})\left[e^{\mathrm{i}\pi(\mu_{1}-\mu_{2})/2}+e^{-\mathrm{i}\pi(\mu_{1}-\mu_{2})/2}\right]\\
\nonumber&&\times\biggl[1+\biggl(\lambda_{1}+\lambda_{2}-\frac{\lambda_{1}^{2}}{3}-\frac{\lambda_{2}^{2}}{3}\biggr)f(-k\tau)+\lambda_{1}\lambda_{2}f^2(-k\tau)+\frac{1}{2}\left(\lambda_{1}^2+\lambda_{2}^{2}\right)g(-k\tau)\biggr]\biggr\}\\
\nonumber&=&\left(\frac{H^2_{\ast}}{2\pi\dot{\sigma}_{\ast}}\right)^{2}(1+k^{2}\tau^{2})\biggl\{1-2\epsilon_{\ast}+\left(6\epsilon_{\ast}-2\eta_{\sigma\sigma\ast}-2\xi_{\ast}s^{2}_{\theta\ast}c_{\theta\ast}-\frac{4}{3}\xi^{2}_{\ast}s_{\theta\ast}^{4}\right)f(-k\tau)\\
&&-\xi^{2}_{\ast}s_{\theta\ast}^{4}\left[f^2(-k\tau)+g(-k\tau)\right]\biggr\},
\end{eqnarray}
\begin{eqnarray}\label{PSstar}
\nonumber\bar{\mathcal{P}}_{\mathcal{S}}&=&\left(\frac{H^2_{\ast}}{2\pi\dot{\sigma}_{\ast}}\right)^{2}(1+k^{2}\tau^{2})(1-2\epsilon_{\ast})\biggl\{\frac{\sin^{2}\Psi_{\ast}+\cos^{2}\Theta_{\ast}}{\cos^{2}(\Theta_{\ast}+\Psi_{\ast})}
+\frac{(6\lambda_{1}-2\lambda_{1}^2)\sin^{2}\Psi_{\ast}+(6\lambda_{2}-2\lambda_{2}^2)\cos^{2}\Theta_{\ast}}{3\cos^{2}(\Theta_{\ast}+\Psi_{\ast})}f(-k\tau)\\
\nonumber&&+\frac{\lambda_{1}^2\sin^{2}\Psi_{\ast}+\lambda_{2}^2\cos^{2}\Theta_{\ast}}{\cos^{2}(\Theta_{\ast}+\Psi_{\ast})}\left[f^2(-k\tau)+g(-k\tau)\right]-\frac{\sin\Psi_{\ast}\cos\Theta_{\ast}}{\cos^{2}(\Theta_{\ast}+\Psi_{\ast})}\sin(\Theta_{\ast}+\Psi_{\ast})\left[e^{\mathrm{i}\pi(\mu_{1}-\mu_{2})/2}+e^{-\mathrm{i}\pi(\mu_{1}-\mu_{2})/2}\right]\\
\nonumber&&\times\biggl[1+\biggl(\lambda_{1}+\lambda_{2}-\frac{\lambda_{1}^{2}}{3}-\frac{\lambda_{2}^{2}}{3}\biggr)f(-k\tau)+\lambda_{1}\lambda_{2}f^2(-k\tau)+\frac{1}{2}\left(\lambda_{1}^2+\lambda_{2}^{2}\right)g(-k\tau)\biggr]\biggr\}\\
\nonumber&=&\left(\frac{H^2_{\ast}}{2\pi\dot{\sigma}_{\ast}}\right)^{2}(1+k^{2}\tau^{2})\biggl\{1-2\epsilon_{\ast}+\left[2\epsilon_{\ast}-2\eta_{ss\ast}+2\xi_{\ast}(1+s_{\theta\ast}^{2})c_{\theta\ast}-\frac{4}{3}\xi_{\ast}^{2}s_{\theta\ast}^{2}c_{\theta\ast}^{2}\right]f(-k\tau)\\
&&+\xi_{\ast}^2\left(1+s_{\theta\ast}^{2}c_{\theta\ast}^{2}\right)\left[f^2(-k\tau)+g(-k\tau)\right]\biggr\},
\end{eqnarray}
\begin{eqnarray}\label{CRSstar}
\nonumber\bar{\mathcal{C}}_{\mathcal{RS}}&=&\left(\frac{H^2_{\ast}}{2\pi\dot{\sigma}_{\ast}}\right)^{2}(1+k^{2}\tau^{2})(1-2\epsilon_{\ast})\biggl\{-\frac{\sin(2\Psi_{\ast})+\sin(2\Theta_{\ast})}{2\cos^{2}(\Theta_{\ast}+\Psi_{\ast})}-\frac{(3\lambda_{1}-\lambda_{1}^2)\sin(2\Psi_{\ast})+(3\lambda_{2}-\lambda_{2}^2)\sin(2\Theta_{\ast})}{3\cos^{2}(\Theta_{\ast}+\Psi_{\ast})}f(-k\tau)\\
\nonumber&&-\frac{\lambda_{1}^2\sin(2\Psi_{\ast})+\lambda_{2}^2\sin(2\Theta_{\ast})}{2\cos^{2}(\Theta_{\ast}+\Psi_{\ast})}\bigl[f(-k\tau)^2+g(-k\tau)\bigr]\\
\nonumber&&+\frac{e^{\mathrm{i}\pi(\mu_{1}-\mu_{2})/2}\sin\Psi_{\ast}\sin\Theta_{\ast}+e^{-\mathrm{i}\pi(\mu_{1}-\mu_{2})/2}\cos\Psi_{\ast}\cos\Theta_{\ast}}{\cos^{2}(\Theta_{\ast}+\Psi_{\ast})}\sin(\Theta_{\ast}+\Psi_{\ast})\biggl[1+\biggl(\lambda_{1}+\lambda_{2}-\frac{\lambda_{1}^{2}}{3}-\frac{\lambda_{2}^{2}}{3}\biggr)f(-k\tau)\\
\nonumber&&+\lambda_{1}\lambda_{2}f^2(-k\tau)+\frac{1}{2}\left(\lambda_{1}^2+\lambda_{2}^{2}\right)g(-k\tau)\biggr]\biggr\}\\
\nonumber&=&\left(\frac{H^2_{\ast}}{2\pi\dot{\sigma}_{\ast}}\right)^{2}(1+k^{2}\tau^{2})\biggl\{\left(2\xi_{\ast} s_{\theta\ast}^{3}-2\eta_{\sigma s\ast}-\frac{4}{3}\xi_{\ast}^{2}s_{\theta\ast}^{3}c_{\theta\ast}\right)f(-k\tau)+\xi_{\ast}^{2}s_{\theta\ast}^{3}c_{\theta\ast}\left[f^2(-k\tau)+g(-k\tau)\right]\\
&&+\frac{\mathrm{i}\pi}{3}\xi_{\ast}^{2}s_{\theta\ast}^{3}c_{\theta\ast}\left(2c_{\theta\ast}^{2}+1\right)\biggr\}.
\end{eqnarray}
Here we have used overbars to avoid confusions with the notations in equations \eqref{PR}, \eqref{CRS} and \eqref{PS}. We have dropped nearly scale-invariant factors $(-k\tau)^{3-\mu_{A}-\mu_{B}}$ as usually done for spectra near the Hubble crossing \cite{Lalak:2007vi,Byrnes:2006fr,Wands:2007bd}. This amounts to discarding the divergent $\ln x$ terms in equations (2.61), (2.62), (2.63) of reference \cite{vandeBruck:2014ata}.

Let us comment the above results. First, the results of \cite{Lalak:2007vi} can be recovered if one switches off terms proportional to $\xi^2$. Second, our final expression depends explicitly on $\tau$ and allows us a precise estimate of the spectra around the time of Hubble crossing. At several e-folds after Hubble crossing, we can safely replace the factor $(1+k^2\tau^2)$ by 1, and the functions $f(-k\tau)$, $g(-k\tau)$ by the numbers $0.7296$ and $1.4672$ respectively. The resulted asymptotic values of spectra will be taken as initial conditions for super-horizon evolution in the next subsection.

Another difference with references \cite{Lalak:2007vi,vandeBruck:2014ata} is that the imaginary part of $\mathcal{C}_{\mathcal{RS}}$, proportional to $(\lambda_{1}-\lambda_{2})\tan(\Theta_{\ast}+\Psi_{\ast})$, is nonzero here although very small. In both references \cite{Lalak:2007vi,vandeBruck:2014ata}, the matrix $\tilde{\mathbf{M}}$ is symmetric and can be diagonalized by matrix \eqref{Rinverse} with $\Psi_{\ast}=-\Theta_{\ast}$, therefore the imaginary part of $\mathcal{C}_{\mathcal{RS}}$ vanishes exactly. The numerical simulations in section \ref{sect-num} are in favor of our nonvanishing result, see figure \ref{fig-Crel}.

\subsection{Super-horizon evolution}\label{subsect-sup}
In the previous subsection, we have studied the evolution of perturbations from the sub-horizon scale to several e-folds after Hubble crossing. In this subsection, taking the asymptotic values of \eqref{PRstar}, \eqref{PSstar}, \eqref{CRSstar} as initial conditions, we will study the evolution of perturbations on the super-horizon scales $k^2\tau^2\ll0$.

As preparation, it is interesting to note that $Q_{\sigma}$ and its time derivative in equation \eqref{pert2} can be eliminated simultaneously with the help of \eqref{pert0}, resulting in
\begin{equation}\label{pert3}
\ddot{\delta s}+3H \dot{\delta s}+\left(\frac{k^2}{a^2}+C_{ss}+\frac{4V_{s}^{2}}{\dot{\sigma}^2}\right)\delta s+\frac{4M_{p}^{2}V_{s}}{\dot{\sigma}^{2}}\frac{k^{2}}{a^{2}}\Phi=0.
\end{equation}
More interestingly, in terms of the curvature perturbation $\mathcal{R}$ and the entropy perturbation $\mathcal{S}$, we can translate equations \eqref{pert0} and \eqref{pert3} into
\begin{eqnarray}
&&\dot{\mathcal{R}}=2(\dot{\theta}+b_{\phi}\dot{\sigma}s_{\theta})\mathcal{S}-\frac{2M_{p}^{2}H}{\dot{\sigma}^{2}}\frac{k^{2}}{a^{2}}\Phi,\label{dR}\\
&&\ddot{\mathcal{S}}+\left(3H-\frac{2\dot{H}}{H}+\frac{2\ddot{\sigma}}{\dot{\sigma}}\right)\dot{\mathcal{S}}+\left(\frac{k^2}{a^2}+C_{ss}+\frac{4V_{s}^{2}}{\dot{\sigma}^2}+6\dot{H}-V_{\sigma\sigma}+b_{\phi}s_{\theta}c_{\theta}V_{\chi}-\frac{V_{s}\dot{\theta}}{\dot{\sigma}}
-\frac{2\dot{\sigma}V_{\sigma}}{M_{p}^{2}H}+\frac{2\dot{H}^{2}}{H^{2}}\right)\mathcal{S}\nonumber\\
&&+\frac{4M_{p}^{2}HV_{s}}{\dot{\sigma}^{3}}\frac{k^{2}}{a^{2}}\Phi=0,\label{dS}
\end{eqnarray}
where equations \eqref{Fried2}, \eqref{sigeom} have been used. In the large-scale limit $k^{2}/(a^{2}H^{2})\rightarrow0$, the $k^2$-terms are negligible, then the entropy perturbation evolves independently and works as a source term for the curvature perturbation. Unlike the canonical situation \cite{Wands:2007bd}, even if the inflation trajectory is straight in field space ($\dot{\theta}=0$), the curvature perturbation is not conserved in the large-scale limit in the presence of entropy perturbation as long as $\xi s_{\theta}\neq0$. This conclusion does not rely on slow-roll conditions, initial conditions or sub-horizon evolution.

For the same reason, we omit $k^2$-terms in equations \eqref{pert0} and \eqref{pert3}, and write down perturbation equations on the super-Hubble scale
\begin{eqnarray}
&&\dot{\sigma}\dot{Q}_{\sigma}+\left(3H+\frac{\dot{H}}{H}\right)\dot{\sigma}Q_{\sigma}+V_{\sigma}Q_{\sigma}+2V_{s}\delta s=0,\label{pertout0}\\
&&\ddot{\delta s}+3H\dot{\delta s}+\left(C_{ss}+\frac{4V_{s}^{2}}{\dot{\sigma}^2}\right)\delta s=0.\label{pertout3}
\end{eqnarray}
Starting from the Hubble-crossing spectra \eqref{PRstar}, \eqref{PSstar}, \eqref{CRSstar}, in this section we will analytically evolve perturbations on the super-Hubble scales according to equations \eqref{pertout0}, \eqref{pertout3} with a precision of $\mathcal{O}\left(\epsilon,\eta,\xi^{2}\right)$.

To the first order of $\xi$, we can neglect the $\ddot{\delta s}$ term in equation \eqref{pertout3} by virtue of the slow-roll conditions. This approximation is supported by the resulted first-order differential equation
\begin{equation}\label{dvs1}
\dot{\delta s}+\frac{1}{3H}\left(C_{ss}+\frac{4V_{s}^{2}}{\dot{\sigma}^2}\right)\delta s+\mathcal{O}\left(\xi^{2}\right)=0,
\end{equation}
which implies $\dot{\delta s}=\mathcal{O}\left(\xi\right)$ because
\begin{equation}
C_{ss}+\frac{4V_{s}^{2}}{\dot{\sigma}^2}=H^{2}\left[3\eta_{ss}-3\xi c_{\theta}(1+s_{\theta}^{2})+\xi^{2}(s_{\theta}^{2}c_{\theta}^{2}+3s_{\theta}^{6}-1)\right]+\mathcal{O}\left(\epsilon^{2},\eta^{2},\xi^{3},\epsilon\eta,\epsilon\xi,\eta\xi\right).
\end{equation}
Differentiating equation \eqref{dvs1} with the help of equation \eqref{dthet}, we get
\begin{equation}\label{ddvs}
\ddot{\delta s}+H\left[\eta_{ss}-\xi c_{\theta}\left(1+s_{\theta}^{2}\right)\right]\dot{\delta s}+2H^2\xi^{2}s_{\theta}^{2}c_{\theta}^{4}\delta s+\mathcal{O}\left(\epsilon^{2},\eta^{2},\xi^{3},\epsilon\eta,\epsilon\xi,\eta\xi\right)=0.
\end{equation}
Now it is clear that $\ddot{\delta s}=\mathcal{O}\left(\xi^2\right)$ as we have expected. We can eliminate the $\ddot{\delta s}$ term in equations \eqref{pertout3} and \eqref{ddvs}, obtaining
\begin{equation}\label{dvs2}
H\left[3-\eta_{ss}+\xi c_{\theta}\left(1+s_{\theta}^{2}\right)\right]\dot{\delta s}+H^2\left[3\eta_{ss}-3\xi c_{\theta}(1+s_{\theta}^{2})-\xi^{2}(1+s_{\theta}^{2}-3s_{\theta}^{4}-s_{\theta}^{6})\right]\delta s+\mathcal{O}\left(\epsilon^{2},\eta^{2},\xi^{3},\epsilon\eta,\epsilon\xi,\eta\xi\right)=0.
\end{equation}

Neglecting corrections of $\mathcal{O}\left(\epsilon^{2},\eta^{2},\xi^{3},\epsilon\eta,\epsilon\xi,\eta\xi\right)$, we can write equations \eqref{pertout0} and \eqref{dvs2} in the form
\begin{equation}\label{dQdvs}
\dot{Q_{\sigma}}\simeq AHQ_{\sigma}+BH\delta s,~~~~\dot{\delta s}\simeq DH\delta s,
\end{equation}
where
\begin{eqnarray}\label{ABD}
\nonumber A&=&2\epsilon-\eta_{\sigma\sigma}-\xi s_{\theta}^{2}c_{\theta}-\frac{2}{3}\xi^{2}s_{\theta}^{2}c_{\theta}^{2},\\
\nonumber B&=&2\xi s_{\theta}^{3}-2\eta_{\sigma s}+\frac{4}{3}\xi^{2}s_{\theta}^{3}c_{\theta},\\
D&=&\xi c_{\theta}(1+s_{\theta}^{2})-\eta_{ss}-\frac{2}{3}\xi^{2}s_{\theta}^{4}.
\end{eqnarray}
Differential equations \eqref{dQdvs} can be solved formally by \eqref{Qvs-dBR}, from which one can write down the power spectra and correlation \eqref{PR-dBR}, \eqref{CRS-dBR}, \eqref{PS-dBR}, see also reference \cite{vandeBruck:2014ata,Avgoustidis:2011em}. Unfortunately, the solution involves double integrals that cannot be performed analytically for the case studied in the next section.
In the main body of this paper, we will take a more efficient approach by writing the solution as
\begin{eqnarray}\label{Qvs}
\nonumber Q_{\sigma}(N)&\simeq&Q_{\sigma\ast}e^{\int^{N}_{N_{\ast}}AdN}+\delta s_{\ast}\left(Ye^{\int^{N}_{N_{\ast}}DdN}-Y_{\ast}e^{\int^{N}_{N_{\ast}}AdN}\right),\\
\delta s(N)&\simeq&\delta s_{\ast}e^{\int^{N}_{N_{\ast}}DdN},
\end{eqnarray}
where the coefficient $Y$ is subject to a differential equation
\begin{equation}\label{dY}
\dot{Y}+Y(D-A)H=BH.
\end{equation}
This differential equation can be integrated explicitly to reproduce \eqref{Qvs-dBR} from \eqref{Qvs}. However, since the explicit solution contains the same double integral, in the next section we will solve equation \eqref{dY} perturbatively. For constant $A$, $B$, $D$, equation \eqref{Qvs} goes back to the solution (86) in reference \cite{Lalak:2007vi}.

Here we should point out that, barring terms of $\mathcal{O}\left(\epsilon^{2},\eta^{2},\xi^{3},\epsilon\eta,\epsilon\xi,\eta\xi\right)$, reference \cite{vandeBruck:2014ata} gives a different value of $D$, here notated as
\begin{equation}\label{D-dBR}
D_{dBR}=\xi c_{\theta}(1+s_{\theta}^{2})-\eta_{ss}-\frac{4}{3}\xi^{2}s_{\theta}^{4}.
\end{equation}
This is not surprising, because in their paper $C_{ss}$ is slightly different from our result \eqref{Csr}.

Under the slow-roll approximation, we can make use of equation \eqref{sigma} to prove the following equation
\begin{equation}
\frac{1}{H}\frac{d}{dt}\ln\frac{H}{\dot{\sigma}}=-2\epsilon+\eta_{\sigma\sigma}+\xi s_{\theta}^{2}c_{\theta}+\frac{2}{3}\xi^{2}s_{\theta}^{2}c_{\theta}^{2}=-A
\end{equation}
and then integrate it,
\begin{equation}\label{Hosig}
\frac{H}{\dot{\sigma}}=\frac{H_{\ast}}{\dot{\sigma}_{\ast}}e^{-\int^{N}_{N_{\ast}}AdN}.
\end{equation}
Recall that $Q_{\sigma}$ and $\delta s$ are related to the curvature and entropy perturbations by \eqref{RS}. Therefore, corresponding to solution \eqref{Qvs}, the power spectra and correlation spectrum on super-Hubble scales are
\begin{eqnarray}
\mathcal{P}_{\mathcal{R}}(N)&\simeq&\mathcal{P}_{\mathcal{R}\ast}+\mathcal{P}_{\mathcal{S}\ast}\left(Ye^{\tilde{\gamma}}-Y_{\ast}\right)^{2}+2\mathrm{Re}(\mathcal{C}_{\mathcal{RS}\ast})\left(Ye^{\tilde{\gamma}}-Y_{\ast}\right),\label{PR}\\
\mathcal{C}_{\mathcal{RS}}(N)&\simeq&\mathcal{C}_{\mathcal{RS}\ast}e^{\tilde{\gamma}}+\mathcal{P}_{\mathcal{S}\ast}e^{\tilde{\gamma}}\left(Ye^{\tilde{\gamma}}-Y_{\ast}\right),\label{CRS}\\
\mathcal{P}_{\mathcal{S}}(N)&\simeq&\mathcal{P}_{\mathcal{S}\ast}e^{2\tilde{\gamma}},\label{PS}
\end{eqnarray}
where $\tilde{\gamma}=\int^{N}_{N_{\ast}}(D-A)dN$. In the next section, we will find analytical expressions for $\tilde{\gamma}$ and $Y$ approximately. What is more, $\mathcal{P}_{\mathcal{R}\ast}$, $\mathcal{P}_{\mathcal{S}\ast}$, $\mathcal{C}_{\mathcal{RS}\ast}$ can be estimated by the asymptotic values of $\bar{\mathcal{P}}_{\mathcal{R}}$, $\bar{\mathcal{P}}_{\mathcal{S}}$, $\bar{\mathcal{C}}_{\mathcal{RS}}$, namely by taking the $k\tau\rightarrow0$ limit of \eqref{PRstar}, \eqref{PSstar}, \eqref{CRSstar}. It is intriguing to note that at large e-folds, equations \eqref{PR}, \eqref{CRS} and \eqref{PS} are all dominated by terms proportional to $\mathcal{P}_{\mathcal{S}\ast}$. By definition of the relative correlation coefficient \eqref{Crel}, $|\tilde{\mathcal{C}}|$ tends to 1 at large $N$. This explains the behavior of $|\tilde{\mathcal{C}}|$ in figure \ref{fig-Crel}.

When the entropy perturbation is absent, the curvature perturbation remains frozen on super-Hubble scales as a consequence of equation \eqref{dR}. In this particular situation, the power spectra of curvature perturbation is given by the single-field result
\begin{equation}\label{Psf}
\mathcal{P}^{\mathrm{sf}}_{\mathcal{R}}(k)\simeq\frac{H^{4}}{4\pi^{2}\dot{\sigma}^{2}}.
\end{equation}
Its Hubble-crossing value $\mathcal{P}^{\mathrm{sf}}_{\mathcal{R}\ast}$ will be utilized to normalize the power spectra and correlation spectrum when comparing with numerical examples.

\section{Case study: $V(\phi,\chi)=U(\phi)+\frac{1}{2}m_{\chi}^2\chi^2$ with $\cos^2\theta\ll1$}\label{sect-case}
In the previous section, we have solved the background and perturbation equations analytically. In the next section, we will do this numerically to evaluate the power spectra and correlation. However, the analytical expressions \eqref{PR}, \eqref{CRS} and \eqref{PS} cannot be compared directly with numerical simulations unless $\tilde{\gamma}$ and $Y$ are determined. Regarding $A$, $B$, $D$ as constants, reference \cite{Lalak:2007vi} worked out $\tilde{\gamma}$ and $Y$ analytically. For time-dependent $A$, $B$ and $D$, this was done by numerical integration in reference \cite{vandeBruck:2014ata}, as we will do in appendix \ref{app-int}.

In the current section, we will calculate $\tilde{\gamma}$ and $Y$ analytically, taking the time dependence of $A$, $B$, $D$ into account. To this end, we have to solve the differential equations in subsection \ref{subsect-traj}. For most models, this is impossible analytically even though $b_{\phi}$ is constant. A key insight is given by equation \eqref{dthet}. From this equation we can see that fixed points of field trajectories should satisfy $|s_{\theta}c_{\theta}^{2}|\ll1$, that is, $|s_{\theta}|\ll1$ or $|c_{\theta}|\ll1$. By power counting we can infer that near $|c_{\theta}|\ll1$ the trajectories are more steep. Therefore, typical solutions of equation \eqref{dthet} can be sorted into three categories: $\theta\simeq0$, $0<\theta<\pi/2$ (unfixed) and $\theta\simeq\pi/2$. As an application of analytical formulae, it is enough for us here to concentrate on field trajectories near $\theta\simeq\pi/2$, or concretely, $c_{\theta}^2\lesssim\xi^2$. In this region, our expression \eqref{ABD} for $A$ and $D$ yields
\begin{equation}\label{D-A}
D-A=\eta_{\sigma\sigma}-\eta_{ss}-2\epsilon-\frac{2}{3}\xi^{2}+3\xi c_{\theta}.
\end{equation}
At the same time, because equations \eqref{phieom}, \eqref{chieom} can be reformed as
\begin{eqnarray}
\nonumber&&3H\dot{\sigma}c_{\theta}+V_{\phi}=b_{\phi}\dot{\sigma}^{2}s_{\theta}^{2},\\
&&(3H+2b_{\phi}\dot{\sigma}c_{\theta})\dot{\sigma}s_{\theta}+e^{-b}V_{\chi}=0
\end{eqnarray}
under the slow-roll approximation, the smallness of $c_{\theta}$ implies
\begin{equation}
\frac{V_{\phi}}{e^{-b}V_{\chi}}=\frac{3c_{\theta}-\xi s_{\theta}^{2}}{(3+2\xi c_{\theta})s_{\theta}}=\mathcal{O}\left(\xi\right).
\end{equation}
This equation can be combined with equations \eqref{Fried1}, \eqref{Fried2}, \eqref{sigeom}, \eqref{dpV} to prove
\begin{eqnarray}
\nonumber\epsilon&=&-\frac{18M_{p}^{4}H^{2}\dot{H}}{18M_{p}^{4}H^{4}}\\
\nonumber&=&\frac{9M_{p}^{2}H^{2}\dot{\sigma}^{2}}{2V^{2}}\\
&\simeq&\frac{M_{p}^{2}e^{-2b}V_{\chi}^{2}}{2V^{2}}.
\end{eqnarray}
Similarly, we have
\begin{equation}
\eta_{\sigma\sigma}\simeq\eta_{\chi\chi}=\frac{M_{p}^{2}e^{-2b}V_{\chi\chi}}{V}.
\end{equation}

Before going to specific models in the next section, we study now the case with a potential of the form $V(\phi,\chi)=U(\phi)+\frac{1}{2}m_{\chi}^2\chi^2$. If we further assume $U/V\ll1$ in accordance with $|c_{\theta}|\ll1$, then this potential satisfies $V_{\chi}^{2}\simeq2VV_{\chi\chi}$ and thus $\epsilon\simeq\eta_{\sigma\sigma}$. This enables us to calculate $\tilde{\gamma}$ and $Y$ for trajectories near $\theta\simeq\pi/2$ analytically but still keeps some generality. For this form of potential, $\eta_{\phi\chi}=0$ exactly by definition \eqref{sr}. Then relation \eqref{eta} can be combined to give
\begin{equation}\label{etasigs}
\eta_{\sigma s}=(\eta_{\sigma\sigma}-\eta_{ss})s_{\theta}c_{\theta},
\end{equation}
and differential equations \eqref{dthet}, \eqref{deta} are reduced to
\begin{eqnarray}\label{dtheteta}
\nonumber\dot{c_{\theta}}&=&Hc_{\theta}\left(\eta_{\sigma\sigma}-\eta_{ss}-\frac{2}{3}\xi^{2}+\xi c_{\theta}\right),\\
\nonumber\dot{\epsilon}&=&2H\epsilon\left(2\epsilon-\eta_{\sigma\sigma}-\xi c_{\theta}\right),\\
\nonumber\dot{\xi}&=&H\xi\left(2\epsilon-\eta_{\sigma\sigma}-\xi c_{\theta}\right),\\
\nonumber\dot{\eta_{\sigma\sigma}}&=&2H\eta_{\sigma\sigma}\left(\epsilon-\xi c_{\theta}\right),\\
\nonumber\dot{\eta_{ss}}&=&2H\epsilon\eta_{ss},\\
\dot{\eta_{\sigma s}}&=&2H\epsilon\eta_{\sigma s}+H\eta_{\sigma s}\eta_{\sigma\sigma}-H\eta_{\sigma s}\eta_{ss} -2H\eta_{ss}\xi s_{\theta}c_{\theta}^{2}-H\eta_{\sigma s}\xi c_{\theta}-\frac{2}{3}\eta_{\sigma s}\xi^{2}.
\end{eqnarray}
The last equation can be derived from the others and relation \eqref{etasigs}. Note that $\alpha_{\sigma\sigma\sigma}$, $\alpha_{\sigma ss}$, $\alpha_{\sigma\sigma s}$ are suppressed by $c_{\theta}$ because $V_{\phi\phi\chi}=V_{\phi\chi\chi}=V_{\chi\chi\chi}=0$ and $s_{\theta}^{2}\simeq1$.

With the reduced equation \eqref{dtheteta}, it is easy to show
\begin{eqnarray}\label{gammat1}
\nonumber\tilde{\gamma}&=&\int^{N}_{N_{\ast}}(D-A)dN\\
\nonumber&=&\int^{N}_{N_{\ast}}\left(\frac{\dot{c_{\theta}}}{Hc_{\theta}}-\frac{\dot{\eta_{\sigma\sigma}}}{H\eta_{\sigma\sigma}}\right)dN\\
&=&\ln\frac{c_{\theta}}{\eta_{\sigma\sigma}}-\ln\frac{c_{\theta\ast}}{\eta_{\sigma\sigma\ast}}.
\end{eqnarray}
However, in case that $c_{\theta}$ is vanishingly small, this expression is unreliable, because higher-order terms such as $2H\xi\left(\epsilon-\eta_{\sigma\sigma}\right)/3$ will dominate the right hand side of the first equation of \eqref{dtheteta}, as evident in equation \eqref{theta2}. Fortunately, under the circumstance $\xi c_{\theta}\ll\epsilon$, the other equations of \eqref{dtheteta} can be readily solved,
\begin{eqnarray}
\nonumber\epsilon&=&\eta_{\sigma\sigma}+\frac{\left(\epsilon_{\ast}-\eta_{\sigma\sigma\ast}\right)H_{\ast}^{4}}{H^{4}},\\
\nonumber\eta_{\sigma\sigma}&=&\frac{\eta_{\sigma\sigma\ast}H_{\ast}^{2}}{H^{2}},\\
\eta_{ss}&=&\frac{\eta_{ss\ast}H_{\ast}^{2}}{H^{2}}.
\end{eqnarray}
Substituting them into equation \eqref{D-A} and remembering that $\xi=\sqrt{2}b_{\phi}M_{p}\sqrt{\epsilon}$, we can integrate out
\begin{eqnarray}\label{gammat2}
\nonumber\tilde{\gamma}&=&\int^{N}_{N_{\ast}}(D-A)dN\\
\nonumber&=&\int^{H}_{H_{\ast}}\left[\eta_{\sigma\sigma}-\eta_{ss}-\left(2+\frac{4}{3}b_{\phi}^{2}M_{p}^{2}\right)\epsilon\right]\frac{dH}{-\epsilon H}\\
&=&\left(2+\frac{4}{3}b_{\phi}^{2}M_{p}^{2}\right)\ln\frac{H}{H_{\ast}}+\frac{1}{2}\left(\frac{\eta_{ss\ast}}{\eta_{\sigma\sigma\ast}}-1\right)\ln\left[1+\frac{\eta_{\sigma\sigma\ast}}{\epsilon_{\ast}}\left(\frac{H^{2}}{H_{\ast}^{2}}-1\right)\right].
\end{eqnarray}
Asymptotic expressions \eqref{gammat1}, \eqref{gammat2} can be combined in the form
\begin{equation}\label{gammat}
\tilde{\gamma}=\ln\left\{\left(\frac{\xi s_{\theta}^{2}c_{\theta}}{\epsilon}\right)^3\frac{\xi c_{\theta}}{\xi_{\ast}c_{\theta\ast}}+\left[1-\left(\frac{\xi s_{\theta}^{2}c_{\theta}}{\epsilon}\right)^3\right]\left(\frac{H}{H_{\ast}}\right)^{\frac{4}{3}b_{\phi}^{2}M_{p}^{2}+\frac{\eta_{ss\ast}}{\eta_{\sigma\sigma\ast}}-2}\right\}-\frac{3}{2}\ln\frac{\eta_{\sigma\sigma}}{\eta_{\sigma\sigma\ast}},
\end{equation}
where for simplicity we have made use of the approximation $\epsilon\simeq\eta_{\sigma\sigma}$.

In order to obtain $Y$, we introduce $\bar{Y}=B/Y$ and rewrite equation \eqref{dY} as
\begin{equation}\label{dYt}
\frac{\dot{\tilde{Y}}}{H}+\tilde{Y}^{2}=\tilde{Y}\left[D-A+\frac{\dot{B}}{HB}\right]=\tilde{Y}\left(-\eta_{ss}-\frac{2}{3}\xi^{2}+2\xi c_{\theta}\right)
\end{equation}
whose leading-order solution ought to be of the form
\begin{equation}
\tilde{Y}=y_{1}\epsilon+y_{2}\eta_{\sigma\sigma}+y_{3}\eta_{ss}+y_{4}\xi^{2}+y_{5}\xi c_{\theta}+y_{6}c_{\theta}^{2}.
\end{equation}
Substituting this form of $\tilde{Y}$ into equation \eqref{dYt} and making use of equations \eqref{dtheteta}, in principle one can solve the resulted polynomial equations to obtain constants $t_{i}$'s. However, we find the solution is trivial, $\tilde{Y}=0$. On the other hand, if we take $\epsilon\simeq\eta_{\sigma\sigma}$ and
\begin{equation}\label{Yt}
\tilde{Y}=2\eta_{\sigma\sigma}-\eta_{ss}-4\epsilon-\frac{2}{3}\xi^{2}+2\xi c_{\theta},
\end{equation}
we will find
\begin{equation}
\frac{\dot{\tilde{Y}}}{H}+\tilde{Y}^{2}-\tilde{Y}\left(-\eta_{ss}-\frac{2}{3}\xi^{2}+2\xi c_{\theta}\right)=2\xi c_{\theta}\left(2\epsilon-\eta_{ss}\right).
\end{equation}
This is the only nontrivial solution that satisfies the most polynomial equations derived from equation \eqref{dYt}. Therefore, in the coming section, we will estimate $Y$ with the following expression
\begin{equation}\label{Y}
Y\simeq\frac{B}{D-A-\left(2\epsilon-\eta_{\sigma\sigma}+\xi s_{\theta}^{2}c_{\theta}\right)}.
\end{equation}

If the value of $D$ is replaced by $D_{dBR}$ in equation \eqref{D-dBR} or reference \cite{vandeBruck:2014ata}, parallel arguments lead to
\begin{eqnarray}
\tilde{\gamma}_{dBR}&=&\tilde{\gamma}-\int^{N}_{N_{\ast}}\frac{2}{3}\xi^{2}dN=\tilde{\gamma}+\frac{4}{3}b_{\phi}^{2}M_{p}^{2}\ln\frac{H}{H_{\ast}},\label{gammat-dBR}\\
Y_{dBR}&\simeq&\frac{B}{D-A-\left(2\epsilon-\eta_{\sigma\sigma}+\frac{1}{3}\xi s_{\theta}^{2}c_{\theta}\right)}\label{Y-dBR}
\end{eqnarray}
instead of \eqref{gammat} and \eqref{Y} respectively.

\section{Comparison with numerical examples}\label{sect-num}
In section \ref{sect-ana}, we have calculated the power spectra and correlation for curvature and entropy perturbations analytically. Although the details are complicated, the logic is quite clear: we solve the coupled system of differential equations \eqref{pert1}, \eqref{pert2} with the initial conditions \eqref{init}, and then evaluate the power spectra and correlation according to definitions \eqref{RR}. The same logic can be also applied to numerical calculations. A numerical approach has been developed in reference \cite{Tsujikawa:2002qx} by virtue that $e_{\sigma}$ and $e_{s}$ are two independent Gaussian random variables. In this approach, one has to run the integration algorithm twice. In the first run, one sets $e_{s}=0$ in the initial conditions \eqref{init} and gets the numerical solutions $\mathcal{R}=\mathcal{R}_{1}$ and $\mathcal{S}=\mathcal{S}_{1}$. In the second run, one chooses $e_{\sigma}=0$ and gets $\mathcal{R}=\mathcal{R}_{2}$ and $\mathcal{S}=\mathcal{S}_{2}$. Then the power spectra and correlation spectrum can be evaluated by \cite{Tsujikawa:2002qx}
\begin{eqnarray}\label{num}
\nonumber\mathcal{P_{R}}&=&\frac{k^3}{2\pi^2}\left(|\mathcal{R}_{1}|^2+|\mathcal{R}_{2}|^2\right),\\
\nonumber\mathcal{P_{S}}&=&\frac{k^3}{2\pi^2}\left(|\mathcal{S}_{1}|^2+|\mathcal{S}_{2}|^2\right),\\
\mathcal{C_{RS}}&=&\frac{k^3}{2\pi^2}\left(\mathcal{R}_{1}^{\dagger}\mathcal{S}_{1}+\mathcal{R}_{2}^{\dagger}\mathcal{S}_{2}\right).
\end{eqnarray}
This numerical method was adopted in references \cite{Lalak:2007vi,vandeBruck:2014ata} to judge their analytical results. We will follow it in this paper. Note in this method the correlation $\mathcal{C_{RS}}$ is not guaranteed to be real, agreeing with definition \eqref{RR}.

As an example, we apply our analytical results and this numerical algorithm to a typical model described by
\begin{eqnarray}\label{model}
b(\phi)=\frac{\alpha\phi}{M_{p}},~~~~V(\phi,\chi)=\frac{1}{2}m_{\phi}^2 \phi^2+\frac{1}{2}m_{\chi}^2 \chi^2.
\end{eqnarray}
It belongs to the subcase we have studied analytically in section \ref{sect-case} and can be utilized to judge our analytical results in previous sections. Concretely, in all figures of this paper, we work out three benchmarks with the following choices of parameters
\begin{equation}\label{param}
\begin{array}{lll}
\mathrm{Left~panels:}&\alpha=-1,&m_{\phi}=m_{\chi};\\
\mathrm{Middle~panels:}&\alpha=1,&m_{\phi}=3m_{\chi};\\
 \mathrm{Right~panels:}&\alpha=-\frac{3}{2},&m_{\phi}=m_{\chi}.
\end{array}
\end{equation}
The left-panel benchmark is exactly the double inflation model with non-canonical kinetic terms in reference \cite{Lalak:2007vi}. The middle-panel benchmark is exactly the linear non-canonical double inflation model with $m_{\chi}/m_{\phi}=1/3$ in reference \cite{vandeBruck:2014ata}.

Under the slow-roll approximation, the background equations of motion \eqref{phieom} and \eqref{chieom} reduce to a couple of first-order differential equations. Therefore, to numerically solve the background equations, we need two boundary conditions. Following reference \cite{Lalak:2007vi}, our algorithm starts at eight e-folds before the Hubble crossing and ends at 60 e-folds after Hubble crossing. At the start point $N=-8$, we set $\phi=0$ in accordance with the canonical initial conditions \eqref{init}. At the end point $N=60$, we set $\epsilon=1$ to make sure the slow-roll condition breaks down as the inflation ends. In reference \cite{vandeBruck:2014ata} the boundary condition is different at the end point, so their figures are quantitatively different from ours, but this does not ruin the judgement here as long as the boundary condition imposed on numerical curves is the same as that on analytical curves.

\begin{figure}
  \centering
  \includegraphics[width=0.33\textwidth]{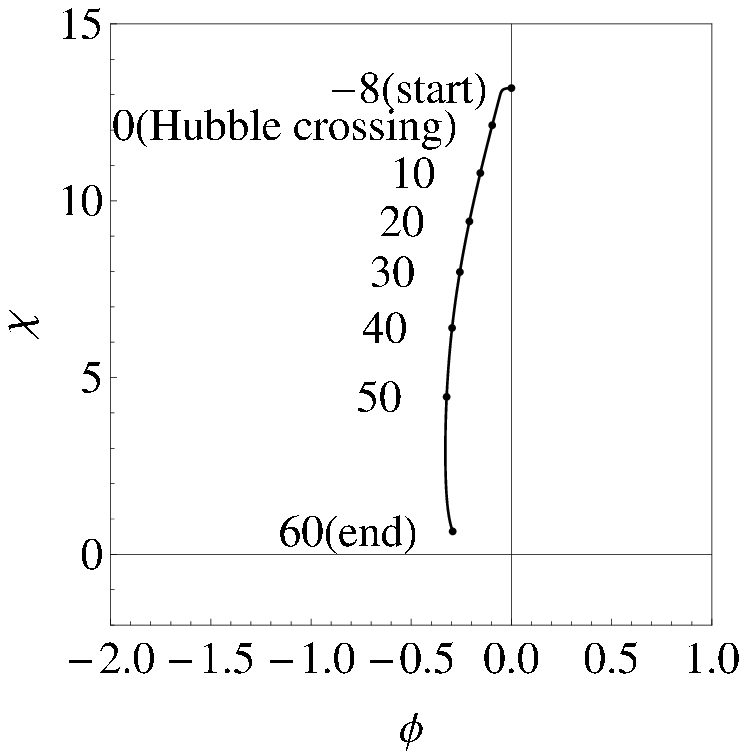}\includegraphics[width=0.33\textwidth]{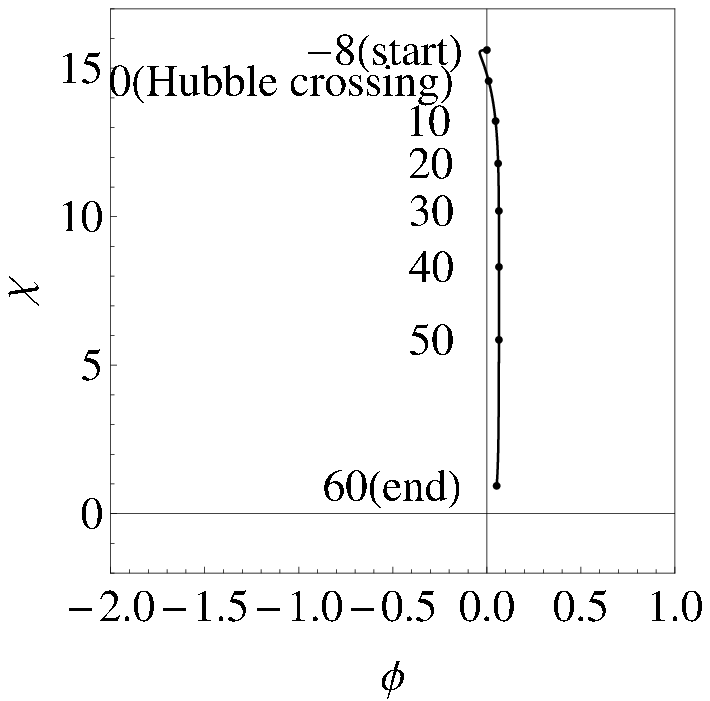}\includegraphics[width=0.33\textwidth]{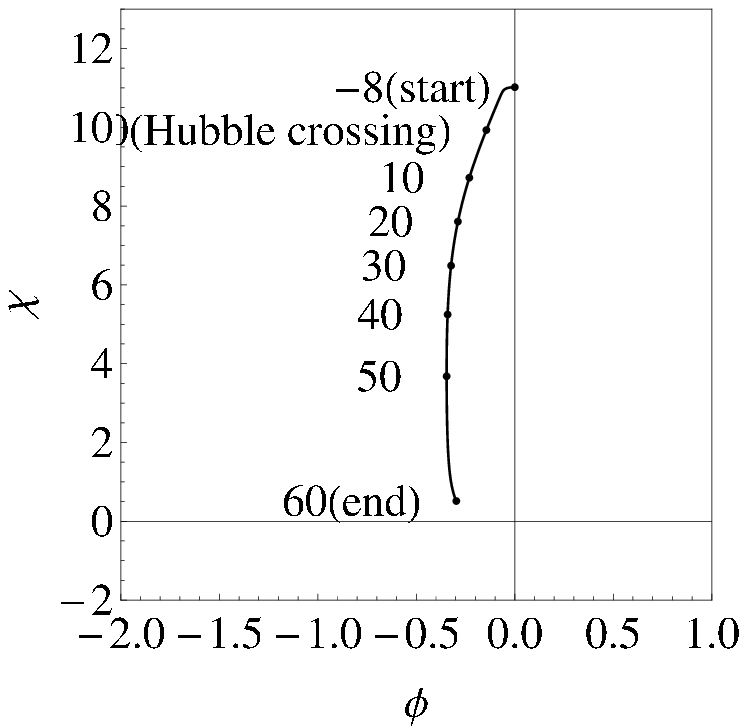}\\
  \caption{(color online). The classical inflationary trajectory for model \eqref{model} with parameters given by \eqref{param}. At the start point $N=-8$, $\phi=0$ as we set. The left panel is the same as figure 1(b) in reference \cite{Lalak:2007vi}, and the middle panel is similar to the top right panel of figure 1 in reference \cite{vandeBruck:2014ata}.}\label{fig-traj}
\end{figure}

\begin{figure}
  \centering
  \includegraphics[width=0.33\textwidth]{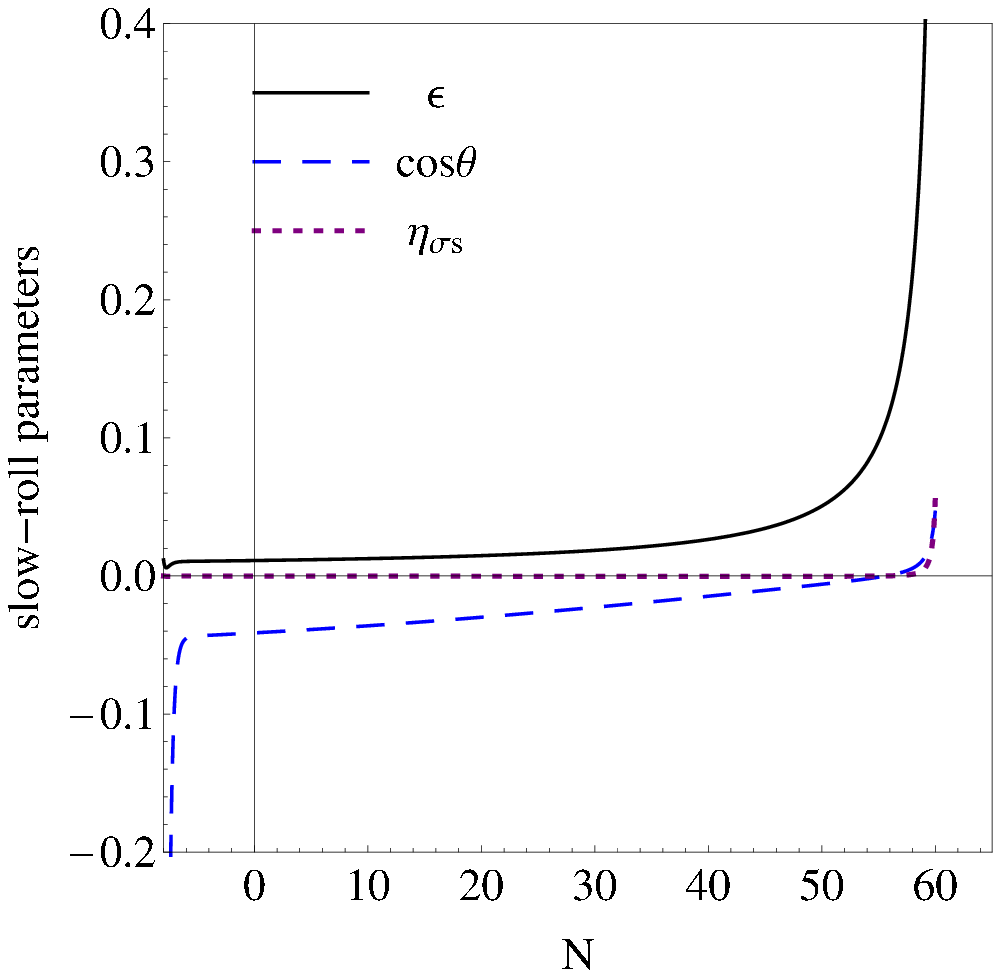}\includegraphics[width=0.33\textwidth]{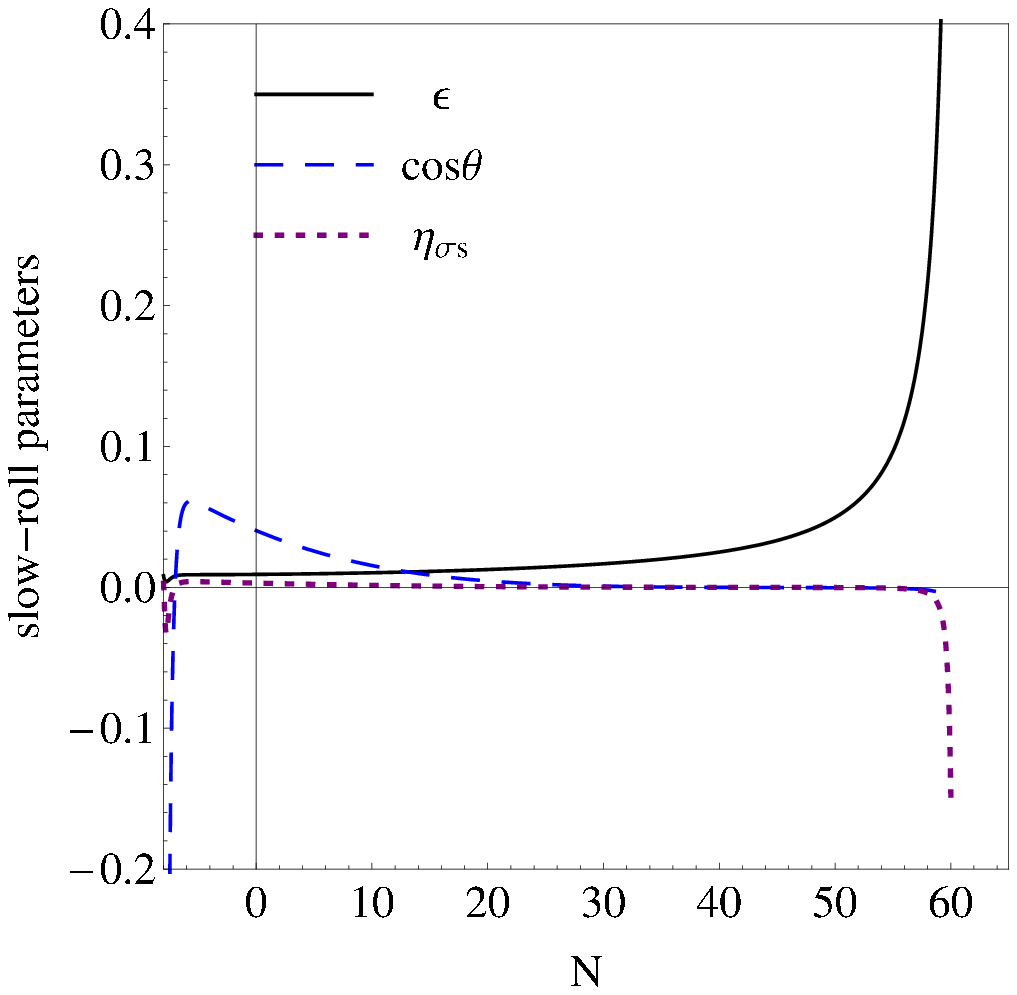}\includegraphics[width=0.33\textwidth]{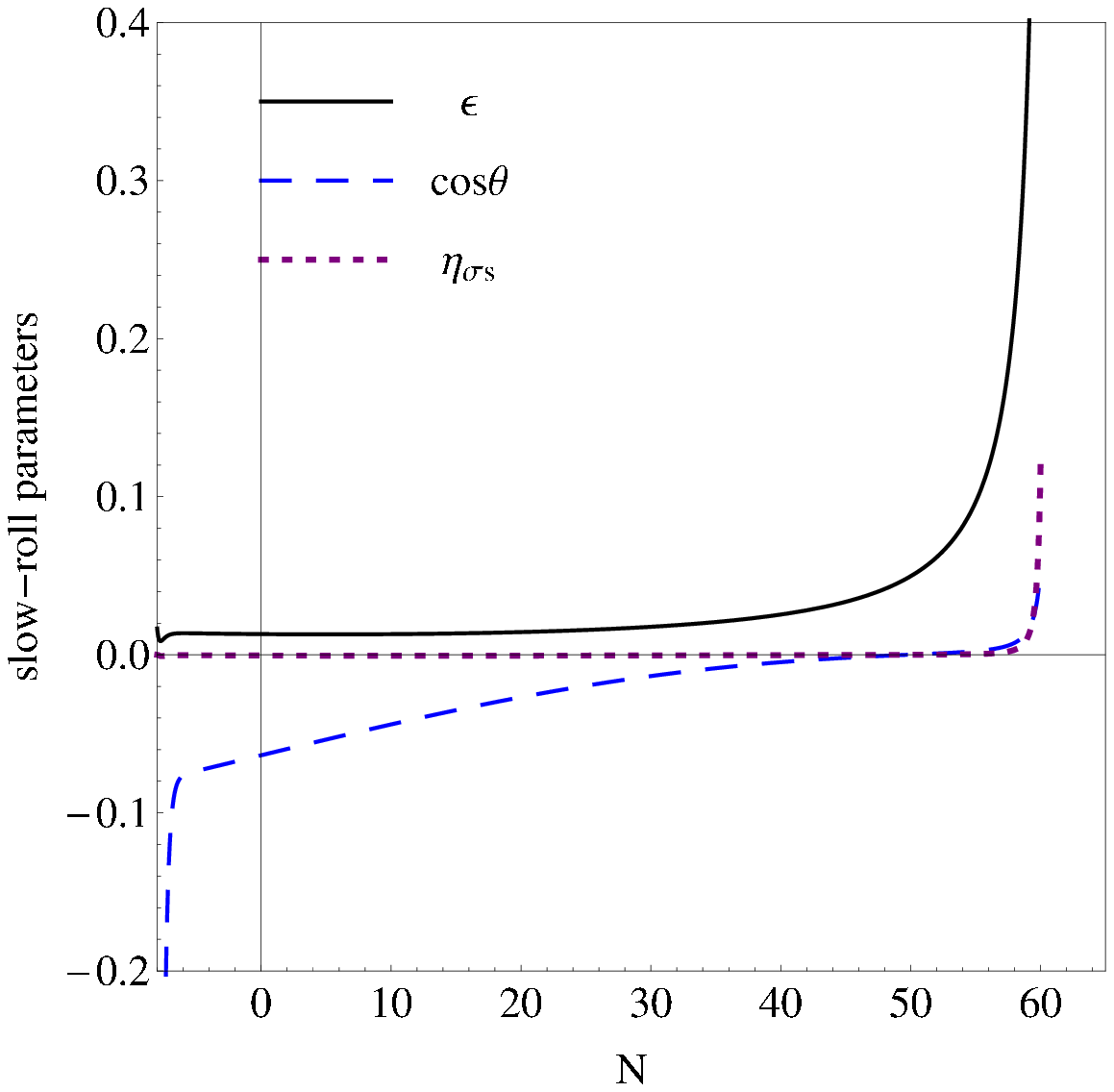}\\
  \caption{(color online). The evolution of $c_{\theta}$ and slow-roll parameters $\epsilon$, $\eta_{\sigma s}$ in model \eqref{model} with parameters given by \eqref{param}. Blue dashed lines, thick solid lines and purple dotted lines depict $c_{\theta}$, $\epsilon$, $\eta_{\sigma s}$ accordingly. At the end point $N=60$, $\epsilon=1$ as we set.}\label{fig-srH}
\end{figure}

Figures \ref{fig-traj} and \ref{fig-srH} correspond to our numerical results of the background evolution. The classical inflationary trajectories in the field space are depicted in figure \ref{fig-traj}. In every panel of the figure, the field $\chi$ decreases monotonically, while field $\phi$ moves from zero to the negative direction and then returns, forming a bent trajectory. The turn point and the exact shape of trajectory are sensitive to model parameters. The evolution of $c_{\theta}$ and slow-roll parameters $\epsilon$, $\eta_{\sigma s}$ are shown in figure \ref{fig-srH} by blue dashed lines, thick solid lines and purple dotted lines accordingly. In all panels of this figure, we can see $\epsilon$ has been kept very small until an instant near the end of inflation $N=60$. In agreement with relation \eqref{etasigs}, $\eta_{\sigma s}$ is strongly suppressed. During inflation, $c_{\theta}$ has passed through the zero line as we have mentioned in section \ref{sect-case}.


\begin{figure}
  \centering
  \includegraphics[width=0.33\textwidth]{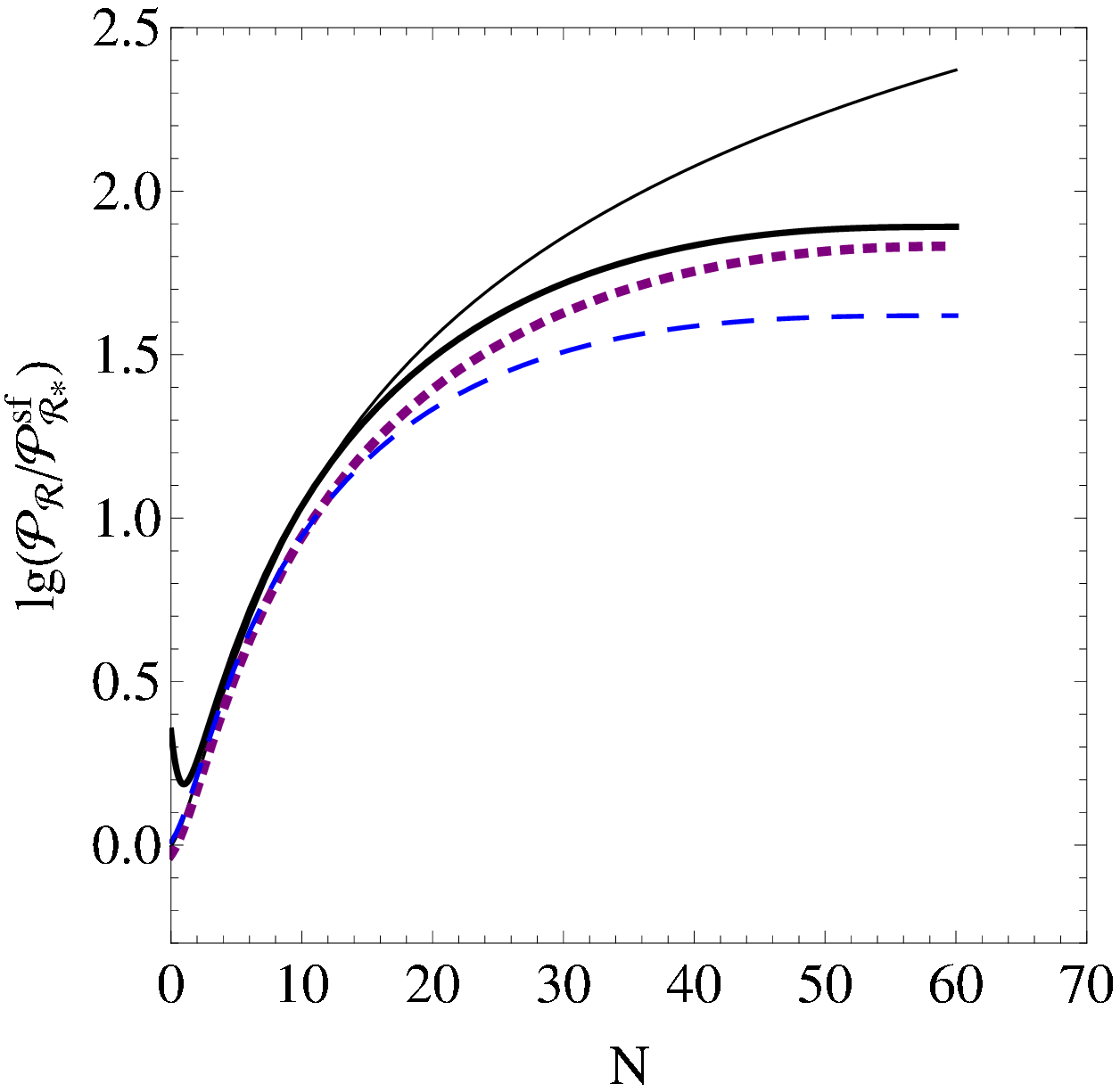}\includegraphics[width=0.33\textwidth]{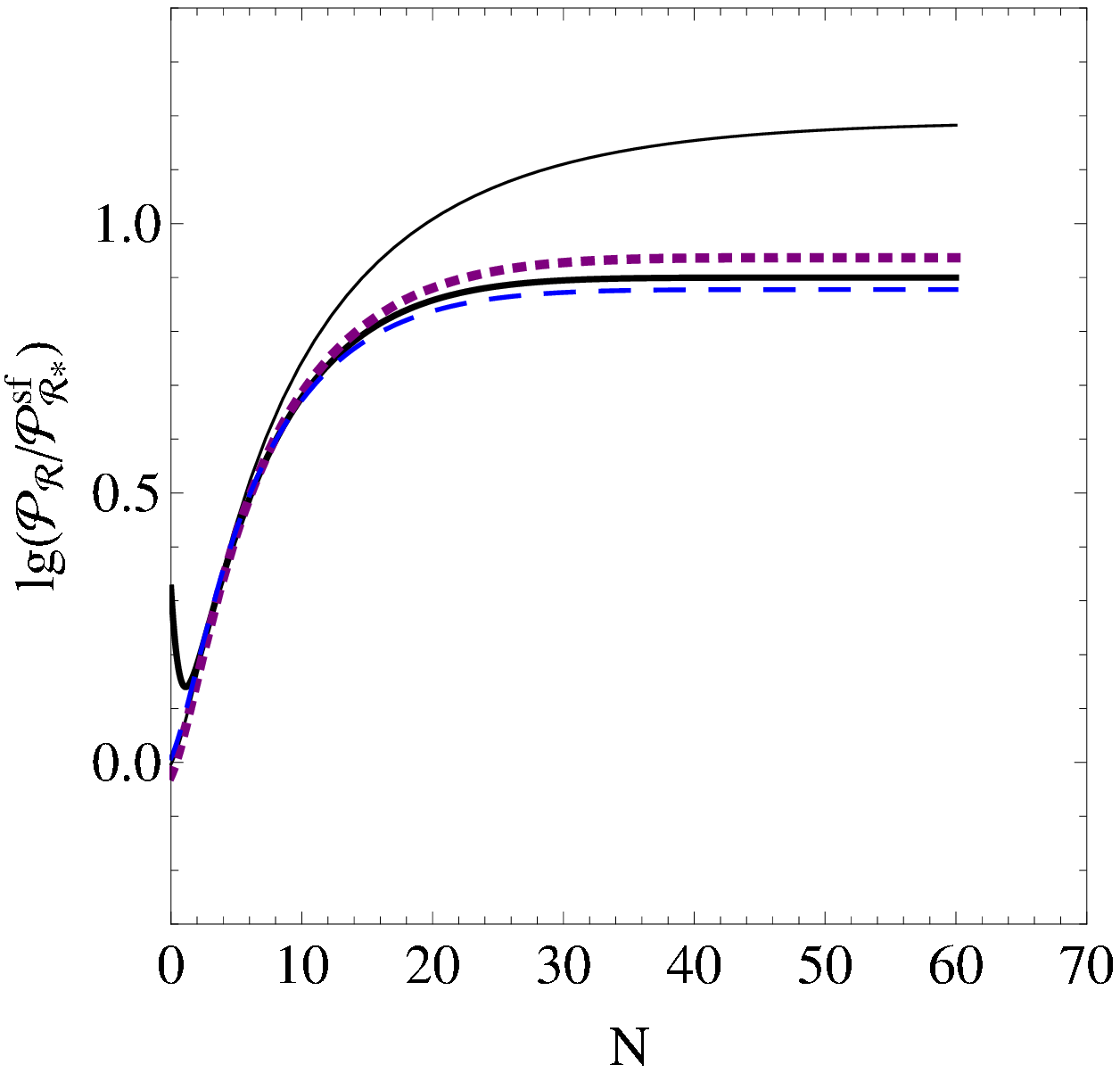}\includegraphics[width=0.33\textwidth]{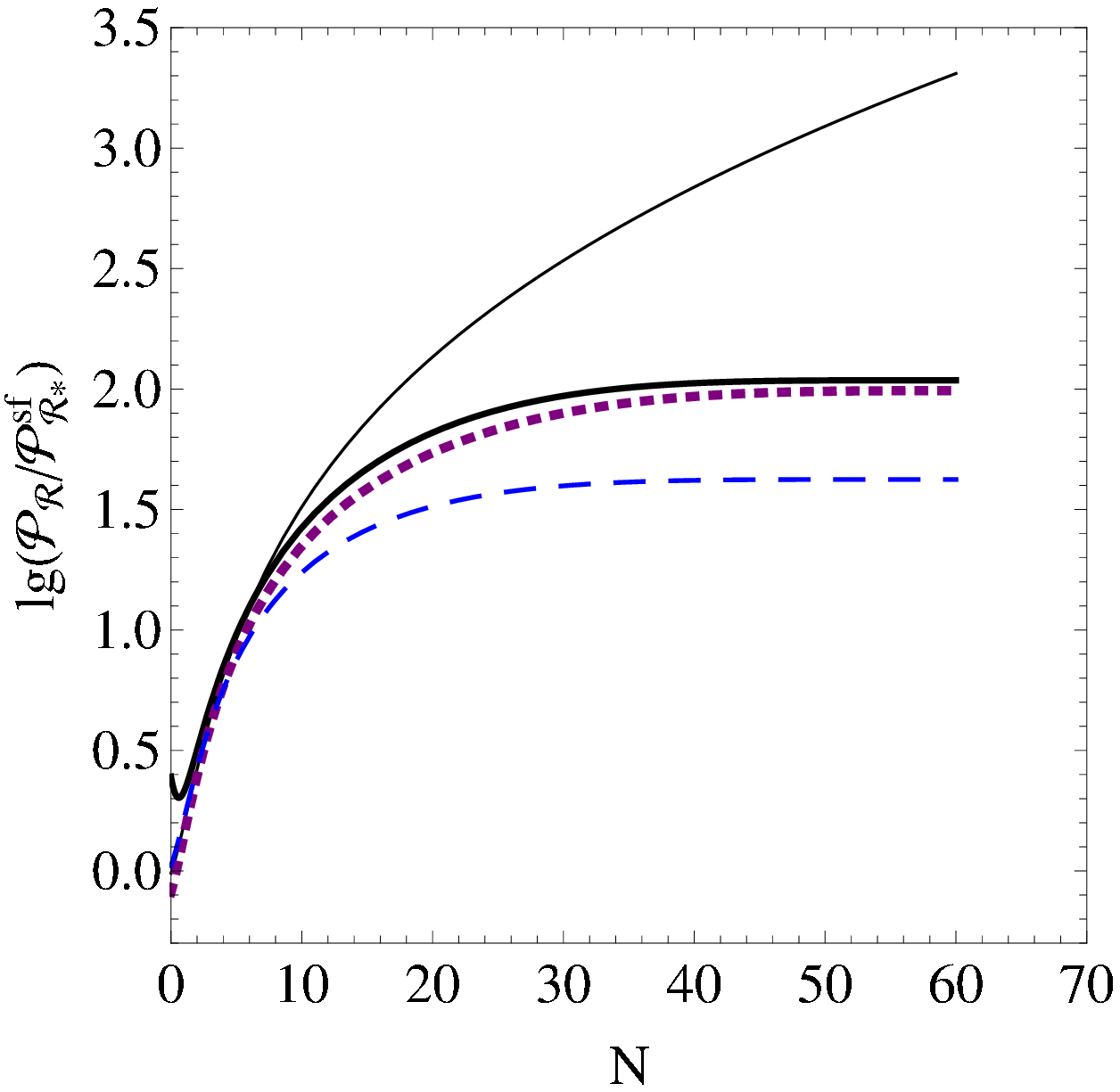}\\
  \caption{(color online). Three different analytical predictions against the numerical result for the power spectrum of curvature perturbation $\mathcal{P_{R}}$ in model \eqref{model} with parameters given by \eqref{param}. Thick solid lines depict the numerical result, and thin solid lines show the analytical prediction of reference \cite{Lalak:2007vi}. Our analytical prediction \eqref{PR} are shown by purple dotted lines, and an analytical approximation for the prediction of reference \cite{vandeBruck:2014ata} is given by blue dashed lines.}\label{fig-PR}
\end{figure}

\begin{figure}
  \centering
  \includegraphics[width=0.33\textwidth]{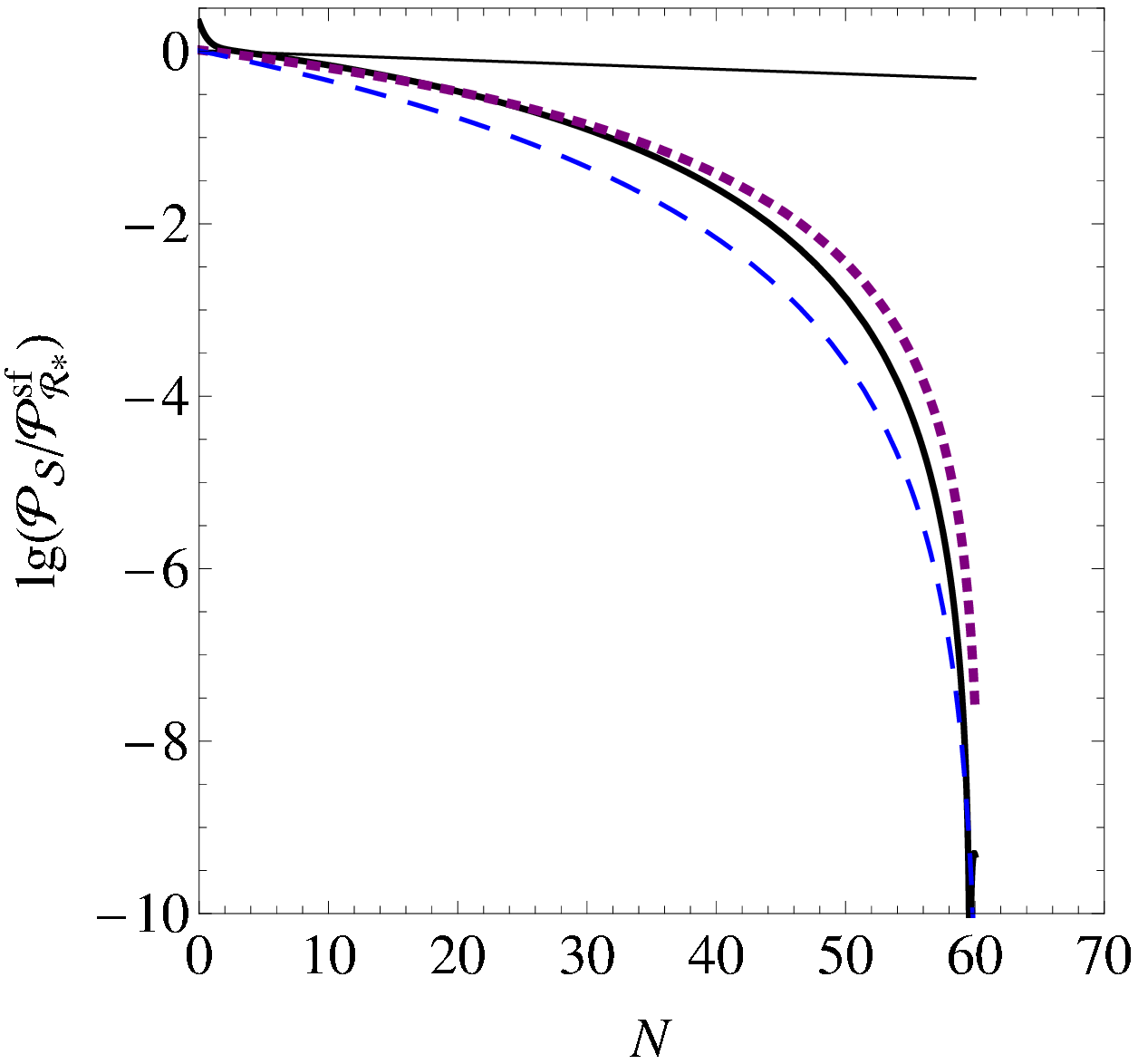}\includegraphics[width=0.33\textwidth]{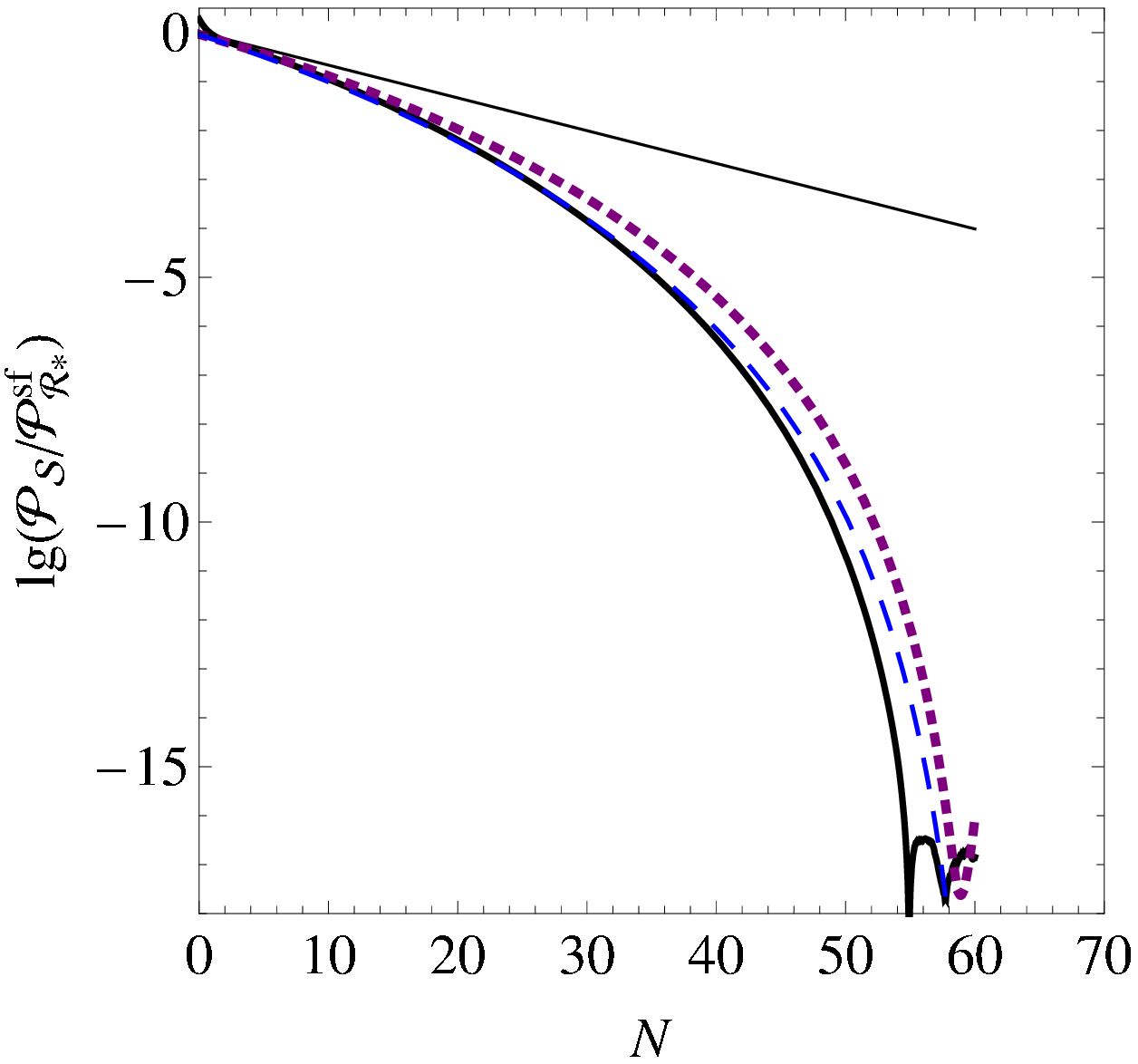}\includegraphics[width=0.33\textwidth]{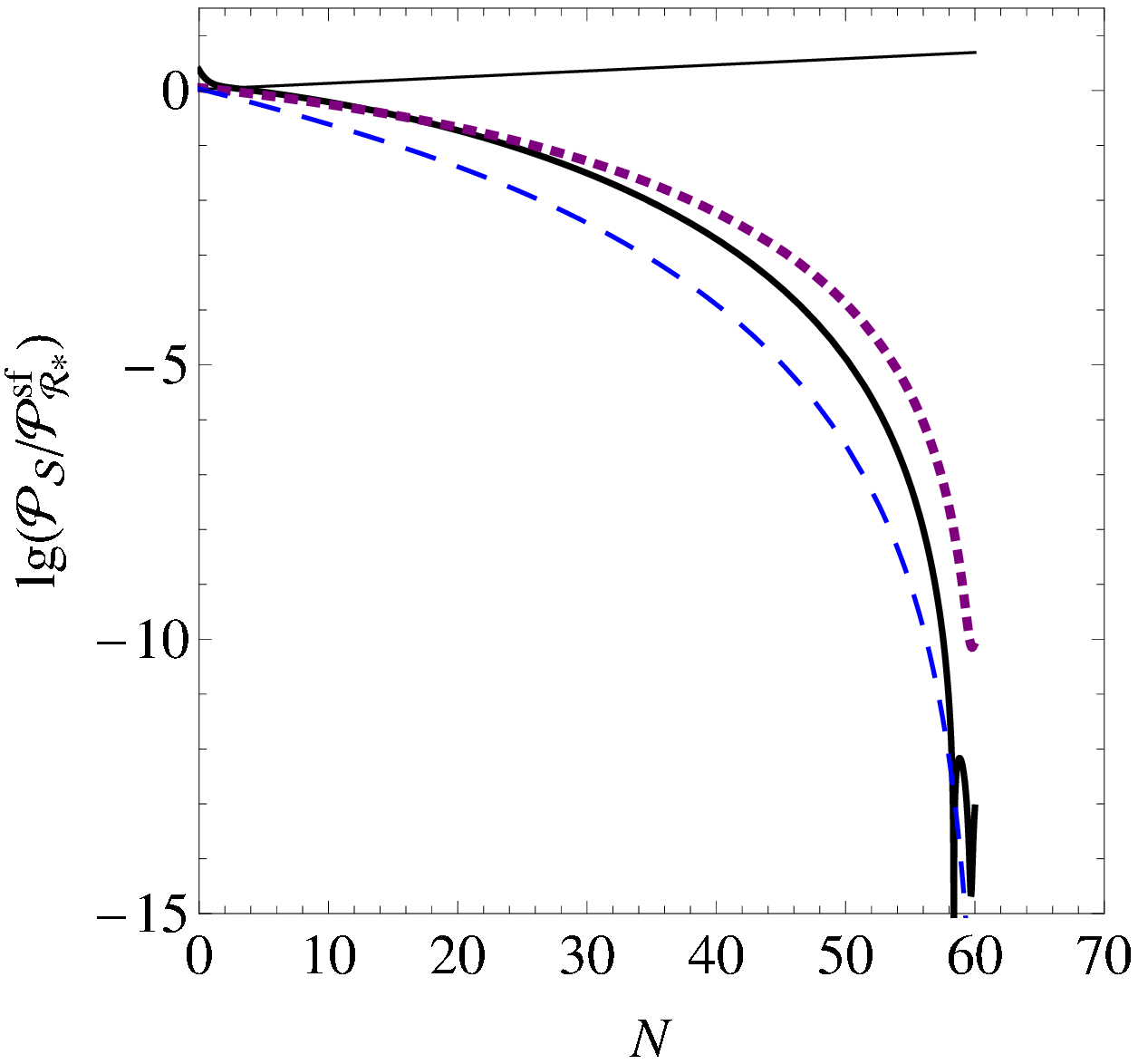}\\
  \caption{(color online). Three different analytical predictions against the numerical result for the power spectrum of entropy perturbation $\mathcal{P_{S}}$ in model \eqref{model} with parameters given by \eqref{param}. Thick solid lines depict the numerical result, and thin solid lines show the analytical prediction of reference \cite{Lalak:2007vi}. Our analytical prediction \eqref{PS} are shown by purple dotted lines, and an analytical approximation for the prediction of reference \cite{vandeBruck:2014ata} is given by blue dashed lines.}\label{fig-PS}
\end{figure}

\begin{figure}
  \centering
  \includegraphics[width=0.33\textwidth]{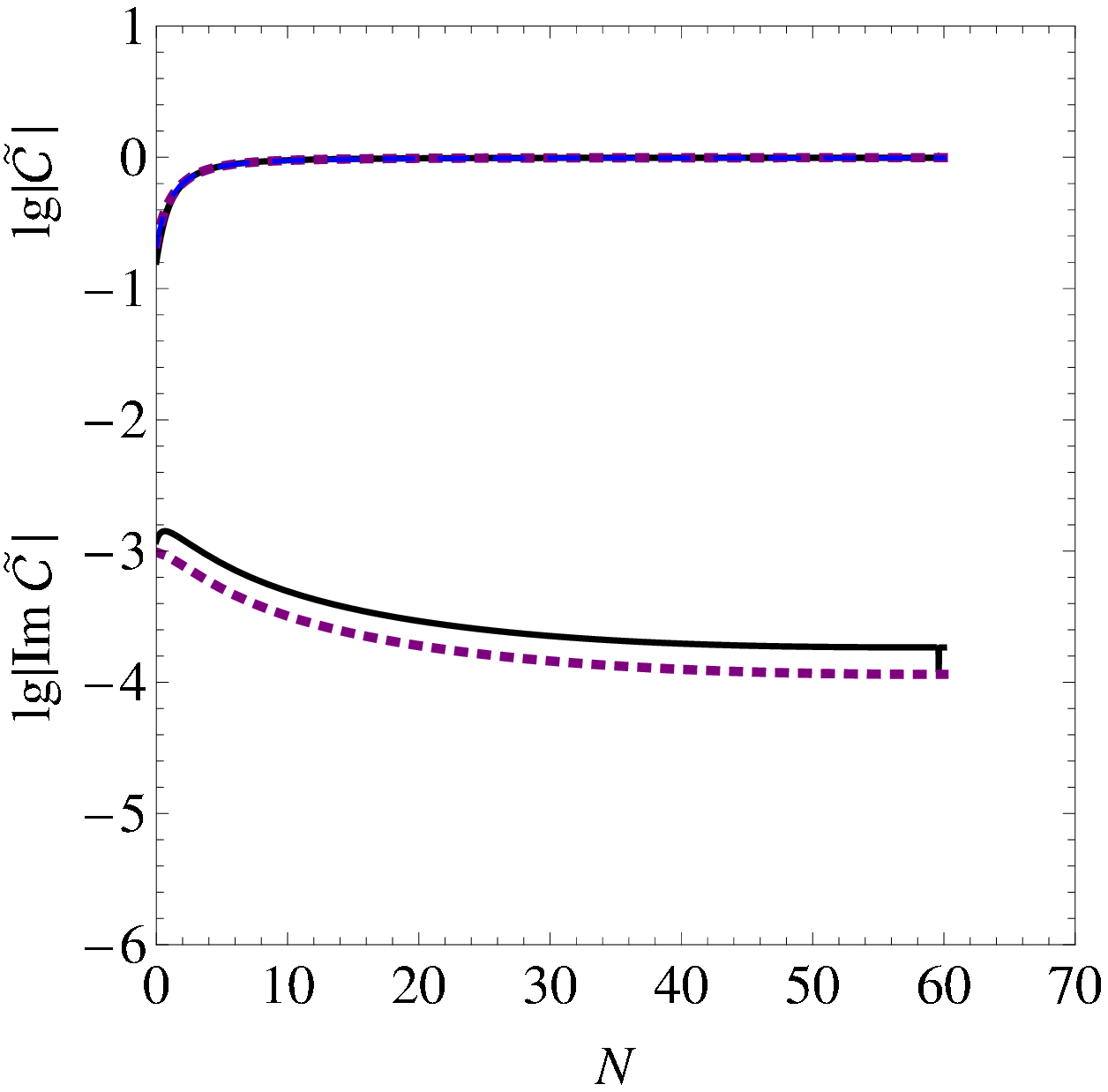}\includegraphics[width=0.33\textwidth]{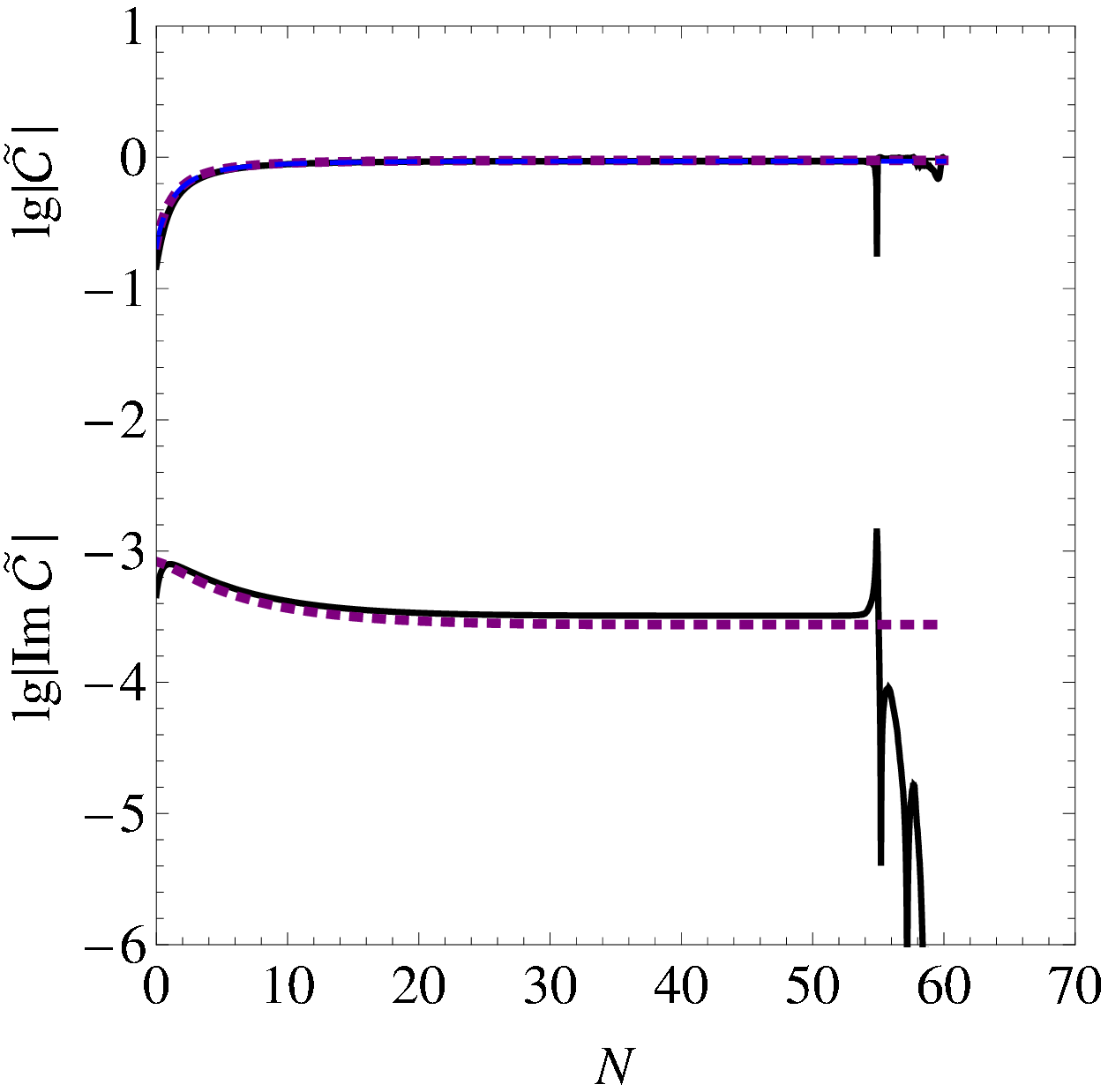}\includegraphics[width=0.33\textwidth]{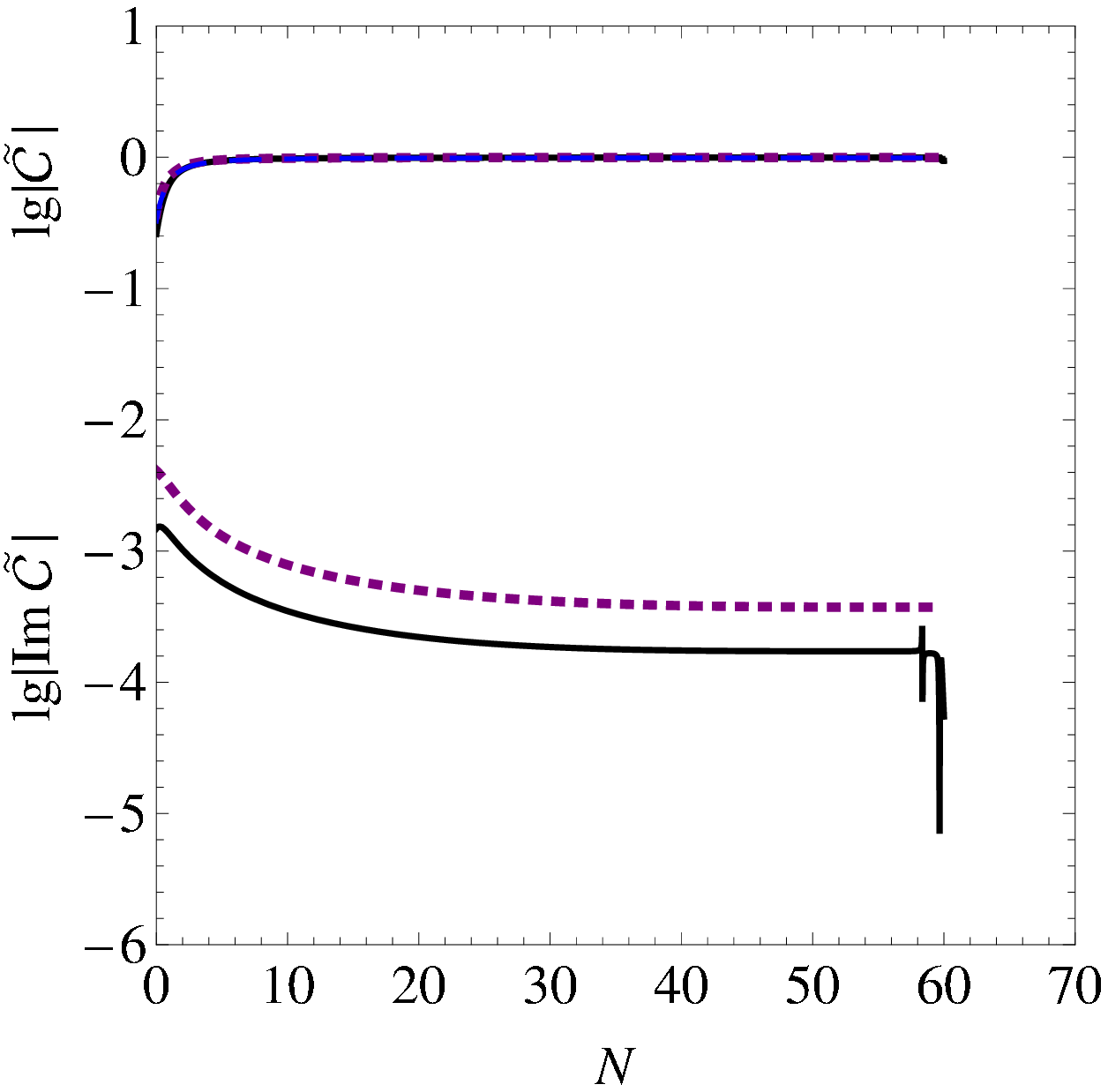}\\
  \caption{(color online). Three different analytical predictions against the numerical result for the relative correlation coefficient $\tilde{\mathcal{C}}$ in model \eqref{model} with parameters given by \eqref{param}. Thick solid lines depict the numerical result, and thin solid lines show the analytical prediction of reference \cite{Lalak:2007vi}. Our analytical prediction \eqref{CRS} are shown by purple dotted lines, and an analytical approximation for the prediction of reference \cite{vandeBruck:2014ata} is given by blue dashed lines. All analytical predictions for $|\tilde{\mathcal{C}}|$ coincide almost perfectly with the numerical result, approaching 1 at large e-folds. Because the imaginary part of $\mathcal{C}_{\mathcal{RS}}$ is nonzero only in our analytical prediction, we plot neither a thin solid line nor a blue dashed line for $\lg|\mathrm{Im~}\tilde{\mathcal{C}}|$.}\label{fig-Crel}
\end{figure}

Figures \ref{fig-PR}, \ref{fig-PS} and \ref{fig-Crel} show analytical predictions for the power spectra and the relative correlation coefficient against numerical results. In these figures, thick solid lines depict results from the full numerical approach \eqref{num}, while thin solid lines reproduce the analytical results of reference \cite{Lalak:2007vi}. The blue dashed lines show our analytical approximation for results of reference \cite{vandeBruck:2014ata}, by replacing $\tilde{\gamma}$, $Y$ in \eqref{PR}, \eqref{CRS} and \eqref{PS} with $\tilde{\gamma}_{dBR}$, $Y_{dBR}$ and switching off the imaginary part in correlation spectrum \eqref{CRSstar}. They fit the numerical results better than thin solid lines. The purple dotted lines depict our analytical results \eqref{PR}, \eqref{CRS} and \eqref{PS} with $\tilde{\gamma}$ and $Y$ given by equations \eqref{gammat}, \eqref{Y}, which in most figures fit the numerical results best.

In figures \ref{fig-PR} and \ref{fig-PS}, the power spectra $\mathcal{P_{R}}$ and $\mathcal{P_{S}}$ are normalized to the Hubble-crossing value of \eqref{Psf}, namely $\mathcal{P}^{\mathrm{sf}}_{\mathcal{R}\ast}\simeq H_{\ast}^{4}/4\pi^{2}\dot{\sigma}_{\ast}^{2}$. As we have discussed in subsection \ref{subsect-pert}, the expression of $C_{ss}$ we obtained in equation \eqref{Csr} is slightly different from the parallel result in reference \cite{vandeBruck:2014ata}. This leads to the difference between blue dashed lines and purple dotted lines. In fact, numerical integration of analytical formulae favors our result \eqref{Csr} in all of the three benchmarks, see appendix \ref{app-int}. In figure \ref{fig-Crel}, we present the magnitude as well as the imaginary part of the relative correlation coefficient $\tilde{\mathcal{C}}$. Because $\mathrm{Im~}\tilde{\mathcal{C}}$ vanishes in references \cite{Lalak:2007vi,vandeBruck:2014ata}, there is neither a thin solid line nor a blue dashed line for $\lg|\mathrm{Im~}\tilde{\mathcal{C}}|$. The magnitude of $\tilde{\mathcal{C}}$ is dominated by its real part, so all analytical curves of $\lg|\tilde{\mathcal{C}}|$ coincide with the numerical result. The large-$N$ behavior of $|\tilde{\mathcal{C}}|$ is also in accordance with our analysis below equation \eqref{PS}.


\section{Comment}\label{sect-com}
In this paper, we analytically computed the primordial power spectra with a precision of $\mathcal{O}\left(\epsilon,\eta,\xi^{2}\right)$ for inflation model \eqref{Sin}. When doing this, we found the perturbation equations \eqref{pertv} cannot be diagonalized through an orthogonal similarity transformation, because a term of order $\xi^{2}$ makes the coefficient matrix \eqref{Mtilde} nonsymmetric. This difficulty can be overcome by introducing a nonorthogonal diagonalization \eqref{diag} and a nonorthogonal basis \eqref{nonorth} at Hubble crossing. To keep the $\mathcal{O}\left(\xi^{2}\right)$ corrections, we expanded the Hankel function \eqref{Hexpand} to the second order. Confronting the difficulty was rewarding: we got a nonzero imaginary part in the correlation spectrum of curvature and entropy perturbations. Future investigation is necessary to extract such a signal from observational data.

The perturbations $Q_{\sigma}$ and $\delta s$ are subject to three differential equations: the constraint \eqref{pert0}, which is a first-order differential equation, and the equations of motion \eqref{pert1}, \eqref{pert2} which are second-order differential equations. Eliminating $Q_{\sigma}$ and $\dot{Q}_{\sigma}$ in equation \eqref{pert2} with \eqref{pert0}, one can obtain the second-order differential equation \eqref{pert3} for $\delta s$. On the super-Hubble scale $k/(aH)\ll1$, the $k^2$-terms can be omitted in these equations. In references \cite{DiMarco:2005nq,Lalak:2007vi,Tsujikawa:2002qx,Bartolo:2001rt}, the super-horizon perturbations are studied with equations \eqref{pert1} and \eqref{pert3}, or more concretely, by truncating the $\ddot{Q}_{\sigma}$ term in equation \eqref{pert1} and $\ddot{\delta s}$ in equation \eqref{pert3}. In subsection \ref{subsect-sup} of this paper, a different approach has been taken: we study the super-horizon perturbations with equation \eqref{pert0} and the truncated form of equation \eqref{pert3}. Unlike the second-order differential equation \eqref{pert1}, equation \eqref{pert0} is a first-order differential equation and hence can be employed without truncation. We have checked but not shown here that incorporating $\mathcal{O}\left(\xi^2\right)$ corrections into the method of references \cite{DiMarco:2005nq,Lalak:2007vi,Tsujikawa:2002qx,Bartolo:2001rt} leads to the same result as equation \eqref{ABD}. Therefore the two methods are equivalent.

One difference of our treatment from that of references \cite{Lalak:2007vi,vandeBruck:2014ata} is that we take $\xi^2$ on the same footing as $\epsilon$. This indeed occurs in several interesting models: the $f(\chi,R)$ generalized gravity \cite{Maeda:1988ab,Nojiri:2003ft,Hwang:2005hb,Ji:2009yw,Saridakis:2016ahq}, the generalized hybrid metric-Palatini gravity \cite{Tamanini:2013ltp}, and the no-scale supergravity inflation \cite{Ellis:2014gxa,Ellis:2014opa}. It will be interesting to apply the formulae and techniques here to observationally relevant subcases of these models.

\begin{acknowledgments}
This work is supported by the National Natural Science Foundation of China (Grant Nos. 11105053 and 91536218).
\end{acknowledgments}

\appendix

\section{Solution of background equations}\label{app-back}
In this appendix, we will solve the system of equations \eqref{3eom} order by order in slow-roll parameters. This will be done by assuming that $\dddot{\sigma}/(H^2\dot{\sigma})$ and $\ddot{\theta}/H^2$ are of $\mathcal{O}\left(\epsilon,\eta,\xi^{2}\right)$ and checking later our solution is consistent with the assumption.

Under the above assumption, we can keep terms of $\mathcal{O}(1)$ in equations \eqref{3eom}
\begin{eqnarray}
\nonumber3\left(3+\frac{V_{\sigma}}{H\dot{\sigma}}\right)+\left(\frac{V_{s}}{H\dot{\sigma}}\right)^2+\mathcal{O}\left(\xi\right)&=&0,\\
-\frac{V_{s}}{H\dot{\sigma}}\left(3+\frac{V_{\sigma}}{H\dot{\sigma}}\right)-\frac{V_{\sigma}}{H\dot{\sigma}}\frac{V_{s}}{H\dot{\sigma}}+\mathcal{O}\left(\xi\right)&=&0
\end{eqnarray}
to find the leading-order result
\begin{eqnarray}
\nonumber\frac{V_{\sigma}}{H\dot{\sigma}}&=&-3+\mathcal{O}\left(\xi\right),\\
\frac{V_{s}}{H\dot{\sigma}}&=&0+\mathcal{O}\left(\xi\right).
\end{eqnarray}
We have dropped the unphysical solution $V_{\sigma}/(H\dot{\sigma})=-4$ because it is in conflict with the condition $\ddot{\sigma}/(H\dot{\sigma})\ll1$ or $\ddot{H}/(H\dot{H})\ll1$.

With the leading-order result taken into consideration, equations \eqref{3eom} accurate to $\mathcal{O}(\xi)$ give
\begin{eqnarray}
\nonumber3\left(3+\frac{V_{\sigma}}{H\dot{\sigma}}\right)-3\xi s_{\theta}^{2}c_{\theta}+\mathcal{O}\left(\epsilon,\eta,\xi^2\right)&=&0,\\
3\left(\frac{V_{s}}{H\dot{\sigma}}+\xi s_{\theta}\right)-3\xi s_{\theta}c_{\theta}^{2}+\mathcal{O}\left(\epsilon,\eta,\xi^2\right)&=&0.
\end{eqnarray}
This leads to the solution of $\mathcal{O}(\xi)$
\begin{eqnarray}
\nonumber\frac{V_{\sigma}}{H\dot{\sigma}}&=&-3+\xi s_{\theta}^{2}c_{\theta}+\mathcal{O}\left(\epsilon,\eta,\xi^2\right),\\
\frac{V_{s}}{H\dot{\sigma}}&=&-\xi s_{\theta}^{3}+\mathcal{O}\left(\epsilon,\eta,\xi^2\right).
\end{eqnarray}
Substituting the solution into equations \eqref{sigeom} and \eqref{theteom}, we can obtain
\begin{eqnarray}
\nonumber\frac{\ddot{\sigma}}{H\dot{\sigma}}&=&-\xi s_{\theta}^{2}c_{\theta}+\mathcal{O}\left(\epsilon,\eta,\xi^2\right),\\
\frac{\dot{\theta}}{H}&=&-\xi s_{\theta}c_{\theta}^{2}+\mathcal{O}\left(\epsilon,\eta,\xi^2\right).
\end{eqnarray}
Remembering that $\xi=b_{\phi}\dot{\sigma}/H$ with a constant $b_{\phi}$ and that that the slow-roll parameters vary slowly during inflation, we can compute the time derivative of the above equations
\begin{eqnarray}
\nonumber\frac{\dddot{\sigma}}{H^{2}\dot{\sigma}}&=&\xi^{2}s_{\theta}^{2}c_{\theta}^{2}(1+c_{\theta}^{2})+\mathcal{O}\left(\epsilon^{2},\eta^{2},\xi^{3},\epsilon\eta,\epsilon\xi,\eta\xi\right),\\
\frac{\ddot{\theta}}{H^2}&=&\xi^{2}s_{\theta}c_{\theta}^{3}\left(2c_{\theta}^{2}-1\right)+\mathcal{O}\left(\epsilon^{2},\eta^{2},\xi^{3},\epsilon\eta,\epsilon\xi,\eta\xi\right).
\end{eqnarray}
This is in agreement with our assumption at the beginning of this section.

With a precision of $\mathcal{O}\left(\epsilon,\eta,\xi^2\right)$, equations \eqref{3eom} give
\begin{eqnarray}
\nonumber\xi^{2}s_{\theta}^{2}c_{\theta}^{2}(1+c_{\theta}^{2})&=&3\left(3+\frac{V_{\sigma}}{H\dot{\sigma}}\right)-\frac{3\dot{H}}{H^2}-\frac{V_{\sigma\sigma}}{H^2}-\xi s_{\theta}^{3}\left(-\xi s_{\theta}^{3}+\xi s_{\theta}\right)-3\xi s_{\theta}^{2}c_{\theta}+\mathcal{O}\left(\epsilon^{2},\eta^{2},\xi^{3},\epsilon\eta,\epsilon\xi,\eta\xi\right),\\
\xi^{2}s_{\theta}c_{\theta}^{3}\left(2c_{\theta}^{2}-1\right)&=&\xi s_{\theta}^{2}c_{\theta}\left(\xi s_{\theta}+\xi s_{\theta}^{3}\right)-\frac{V_{\sigma s}}{H^2}+\left(\xi c_{\theta}+3-\xi s_{\theta}^{2}c_{\theta}\right)\left(\frac{V_{s}}{H\dot{\sigma}}+\xi s_{\theta}\right)-3\xi s_{\theta}c_{\theta}^{2}+\mathcal{O}\left(\epsilon^{2},\eta^{2},\xi^{3},\epsilon\eta,\epsilon\xi,\eta\xi\right),
\end{eqnarray}
whose solution is
\begin{eqnarray}
\nonumber\frac{V_{\sigma}}{H\dot{\sigma}}&=&-3-\epsilon+\eta_{\sigma\sigma}+\xi s_{\theta}^{2}c_{\theta}+\frac{2}{3}\xi^{2}s_{\theta}^{2}c_{\theta}^{2}+\mathcal{O}\left(\epsilon^{2},\eta^{2},\xi^{3},\epsilon\eta,\epsilon\xi,\eta\xi\right),\\
\frac{V_{s}}{H\dot{\sigma}}&=&\eta_{\sigma s}-\xi s_{\theta}^{3}-\frac{2}{3}\xi^{2}s_{\theta}^{3}c_{\theta}+\mathcal{O}\left(\epsilon^{2},\eta^{2},\xi^{3},\epsilon\eta,\epsilon\xi,\eta\xi\right).
\end{eqnarray}
Substituted into equations \eqref{sigeom} and \eqref{theteom}, it yields
\begin{eqnarray}
&&\frac{\ddot{\sigma}}{H\dot{\sigma}}=\epsilon-\eta_{\sigma\sigma}-\xi s_{\theta}^{2}c_{\theta}-\frac{2}{3}\xi^{2}s_{\theta}^{2}c_{\theta}^{2}+\mathcal{O}\left(\epsilon^{2},\eta^{2},\xi^{3},\epsilon\eta,\epsilon\xi,\eta\xi\right),\label{sigma}\\
&&\frac{\dot{\theta}}{H}=-\eta_{\sigma s}-\xi s_{\theta}c_{\theta}^{2}+\frac{2}{3}\xi^{2}s_{\theta}^{3}c_{\theta}+\mathcal{O}\left(\epsilon^{2},\eta^{2},\xi^{3},\epsilon\eta,\epsilon\xi,\eta\xi\right).\label{theta}
\end{eqnarray}

Repeating the same procedure, we can find out terms proportional to $\xi^{3}$, $\epsilon\xi$ and $\eta\xi$ to get
\begin{eqnarray}
&&\frac{\ddot{\sigma}}{H\dot{\sigma}}=\epsilon-\eta_{\sigma\sigma}-\xi s_{\theta}^{2}c_{\theta}-\frac{2}{3}\xi^{2}s_{\theta}^{2}c_{\theta}^{2}-\frac{4}{3}\xi\eta_{\sigma s}s_{\theta}c_{\theta}^{2}+\frac{4}{3}\xi\left(\epsilon-\eta_{\sigma\sigma}\right)s_{\theta}^{2}c_{\theta}-\frac{4}{9}\xi^{3}s_{\theta}^{2}c_{\theta}^{3}+
\mathcal{O}\left(\epsilon^{2},\eta^{2},\xi^{4},\epsilon\eta,\epsilon\xi^{2},\eta\xi^{2}\right),\label{sigma2}\\
&&\frac{\dot{\theta}}{H}=-\eta_{\sigma s}-\xi s_{\theta}c_{\theta}^{2}+\frac{2}{3}\xi^{2}s_{\theta}^{3}c_{\theta}-\frac{2}{3}\xi\left(\epsilon-\eta_{\sigma\sigma}\right)s_{\theta}^{3}-\frac{2}{3}\xi\eta_{ss} s_{\theta}c_{\theta}^{2}+\frac{4}{9}\xi^{3}s_{\theta}^{3}c_{\theta}^{2}
+\mathcal{O}\left(\epsilon^{2},\eta^{2},\xi^{4},\epsilon\eta,\epsilon\xi^{2},\eta\xi^{2}\right).\label{theta2}
\end{eqnarray}

\section{Diagonalization of $\tilde{\mathbf{M}}$}\label{app-diag}
If $(a-d)^2+4bc>0$, an arbitrary matrix of two dimensions
\begin{equation}\label{abcd}
\tilde{\mathbf{M}}=\left(\begin{array}{cc}
 a & b \\
 c & d \\
\end{array}\right)
\end{equation}
can be always diagonalized by transformation \eqref{diag} with $\tilde{\mathbf{R}}^{-1}_{\ast}$ given by \eqref{Rinverse} and
\begin{eqnarray}
\nonumber&&\tilde{\lambda}_{1}=\frac{1}{2} \left[a+d-\sqrt{(a-d)^2+4bc}\right],\\
&&\tilde{\lambda}_{2}=\frac{1}{2} \left[a+d+\sqrt{(a-d)^2+4bc}\right].
\end{eqnarray}

Specifically, when $\tilde{\mathbf{M}}$ takes the form \eqref{Mtilde}, its eigenvalues are
\begin{eqnarray}\label{eigen}
\nonumber\tilde{\lambda}_{1}&\simeq& \left(1+3 s_{\theta}^{2}\right)^{-1/2}\left[ \xi ^2 \left(c_{\theta}+s_{\theta}^{2}c_{\theta}\right)-6 \eta_{\sigma s} s^3_{\theta}+ \left(\frac{3}{2} \eta_{ ss }-\frac{3}{2} \eta_{\sigma\sigma }+3\epsilon\right)\left(2c_ {\theta}^{3}-3c_ {\theta}\right)\right]\\
\nonumber&&+\frac{3}{2} \xi  \left(\sqrt{1+3 s_{\theta}^{2}}- c_ {\theta}\right)+\frac{3}{2} \eta_{ss}+\frac{3}{2}  \eta _{\sigma\sigma} +\frac{1}{2} \xi ^2 c^2_{\theta}-3 \epsilon +\mathcal{O}\left(\epsilon^{2},\eta^{2},\xi^{3},\epsilon\eta,\epsilon\xi,\eta\xi\right),\\
\nonumber\tilde{\lambda}_{2}&\simeq&- \left(1+3 s_{\theta}^{2}\right)^{-1/2}\left[ \xi ^2 \left(c_{\theta}+s_{\theta}^{2}c_{\theta}\right)-6 \eta_{\sigma s} s^3_{\theta}+ \left(\frac{3}{2} \eta_{ ss }-\frac{3}{2} \eta_{\sigma\sigma }+3\epsilon\right)\left(2c_ {\theta}^{3}-3c_ {\theta}\right)\right]\\
&&-\frac{3}{2} \xi  \left(\sqrt{1+3 s_{\theta}^{2}}+ c_ {\theta}\right)+\frac{3}{2} \eta_{ss}+\frac{3}{2}  \eta _{\sigma\sigma} +\frac{1}{2} \xi ^2 c^2_{\theta}-3 \epsilon +\mathcal{O}\left(\epsilon^{2},\eta^{2},\xi^{3},\epsilon\eta,\epsilon\xi,\eta\xi\right),
\end{eqnarray}
and one obtains the following combinations
\begin{eqnarray}
\nonumber\tan(\Theta+\Psi)&=&\frac{2\xi s^{3}_{\theta}c^{3}_{\theta}}{3\sqrt{1+3s_{\theta}^{2}}}+\mathcal{O}\left(\epsilon,\eta,\xi^{2}\right),\\
\nonumber\frac{\cos(\Theta-\Psi)}{\cos(\Theta+\Psi)}&=&-\frac{2c_{\theta}^{3}-3c_{\theta}}{\sqrt{1+3s^2_{\theta}}}-\frac{4 s^3_{\theta}\left[\eta_{\sigma s}\left(6c_{\theta}^{3}-9c_{\theta}\right)+s^3_{\theta}\left(3\eta_{ss}-3\eta_{\sigma\sigma}-2\xi^2+6\epsilon\right)\right]}{3\xi\left(1+3s^2_{\theta}\right)^{3/2}}+\mathcal{O}\left(\epsilon,\eta,\xi^{2}\right),\\
\frac{\sin(\Theta-\Psi)}{\cos(\Theta+\Psi)}&=&-\frac{2s^3_{\theta}}{\sqrt{1+3s^2_{\theta}}}+\frac{(4c^3_{\theta}-6c_{\theta})\left[\eta_{\sigma s}\left(6c_{\theta}^{3}-9c_{\theta}\right)+s^3_{\theta}\left(3\eta_{ss}-3\eta_{\sigma\sigma}-2\xi^2+6\epsilon\right)\right]}{3\xi\left(1+3s^2_{\theta}\right)^{3/2}}+\mathcal{O}\left(\epsilon,\eta,\xi^{2}\right)
\end{eqnarray}
whose higher-order expressions are useful for deriving equations \eqref{PRstar}, \eqref{PSstar} and \eqref{CRSstar}.

\section{Expansion of the Hankel function}\label{app-Hankel}
In terms of the gamma function $\Gamma(\mu)$ and Bessel functions of the first kind $J_{\mu}(x)$, the Hankel function $H_{\mu}^{(1)}(x)$ can be expressed as equations \eqref{HJ} and expanded as
\begin{eqnarray}
\nonumber H_{\mu}^{(1)}(x)&=&\frac{J_{-\mu}(x)-\mathrm{e}^{-\mathrm{i}\mu\pi}J_{\mu}(x)}{\mathrm{i}\sin(\mu\pi)}\\
\nonumber&=&\frac{1}{\mathrm{i}\sin(\mu\pi)}\left[\sum^{\infty}_{m=0}\frac{(-1)^{m}}{m!\Gamma(m-\mu+1)}\left(\frac{x}{2}\right)^{2m-\mu} -\sum^{\infty}_{m=0}\frac{(-1)^{m}}{m!\Gamma(m+\mu+1)}\left(\frac{x}{2}\right)^{2m+\mu} \mathrm{e}^{-\mathrm{i}\mu\pi}\right]\\
\nonumber&\simeq&\frac{1}{\mathrm{i}\sin(\mu\pi)}\left(\frac{2}{x}\right)^{\mu}\left[\frac{1}{\Gamma(1-\mu)}-\frac{1}{\Gamma(2-\mu)}\left(\frac{x}{2}\right)^{2} \right]\\
\nonumber&=&\frac{1}{\mathrm{i}\pi}\left(\frac{2}{x}\right)^{\mu}\Gamma(\mu)\left[1-\frac{\Gamma(1-\mu)}{\Gamma(2-\mu)}\left(\frac{x}{2}\right)^{2} \right]\\
&=&\frac{1}{\mathrm{i}\pi}\left(\frac{2}{x}\right)^{\mu}\Gamma(\mu)\left[1-\frac{1}{1-\mu}\left(\frac{x}{2}\right)^{2}\right]
\end{eqnarray}
for $\mu\simeq3/2$ and $x\ll1$, where we have made use of the equality
\begin{equation}
\frac{\pi}{\sin({\mu\pi})}=\Gamma(\mu)\Gamma(1-\mu).
\end{equation}
Apparently, when $x\rightarrow0$, the Hankel function $H_{\mu}^{(1)}(x)$ is divergent as $x^{-\mu}$. In this paper, we renormalize $H_{\mu}^{(1)}(x)$ as
\begin{equation}
\tilde{H}_{\mu}^{(1)}(x)\equiv x^{\mu}H_{\mu}^{(1)}(x)
\end{equation}
and define
\begin{equation}
f(x)=\left.\frac{1}{\tilde{H}_{3/2}^{(1)}(x)}\frac{d\tilde{H}_{\mu}^{(1)}(x)}{d\mu}\right|_{\mu=3/2},~~~~g(x)=\left.\frac{1}{\tilde{H}_{3/2}^{(1)}(x)}\frac{\mathrm{d^2}\tilde{H}_{\mu}^{(1)}(x)}{d\mu^2}\right|_{\mu=3/2}.
\end{equation}
Then $\tilde{H}_{\mu_{A}}^{(1)}(x)$ can be expanded to $\mathcal{O}\left(\lambda_{A}^{2}\right)$ as \eqref{Hexpand}.

With the help of polygamma function, we can get the numerical values of $f(x)$ and $g(x)$. The polygamma function of order $m$ is defined as
\begin{equation}
\psi^{(m)}(x)\equiv\frac{d^{m+1}}{dx^{m+1}}\ln\Gamma(x),
\end{equation}
which can be calculated by the integrations
\begin{equation}
\psi^{(0)}(x)=-\gamma+\int^1_{0}\frac{1-t^{x-1}}{1-t}dt,~~~~\psi^{(m)}(x)=-\int^1_{0}\frac{t^{x-1}}{1-t}\ln^mtdt
\end{equation}
if $x>0$ and $m>0$. From these formulae, we can derive
\begin{equation}
\psi^{(0)}\left(\frac{3}{2}\right)=2-2\ln2-\gamma,~~~~\psi^{(1)}\left(\frac{3}{2}\right)=\frac{\pi^2}{2}-4.
\end{equation}
Using the above definitions and equations, we can compute $f(x)$ and $g(x)$ to $\mathcal{O}(x^2)$ as
\begin{eqnarray}
\nonumber f(x)&=&\left.\frac{d\ln\tilde{H}_{\mu}^{(1)}(x)}{d\mu}\right|_{\mu=3/2}\\
\nonumber&=&\ln2+\left.\frac{d\ln\Gamma(x)}{d\mu}\right|_{\mu=3/2}-\left.\frac{1}{(1-\mu)^2}\left(\frac{x}{2}\right)^{2}\left[1-\frac{1}{1-\mu}\left(\frac{x}{2}\right)^{2}\right]^{-1}\right|_{\mu=3/2}\\
\nonumber&=&\ln2+\psi^{(0)}\left(\frac{3}{2}\right)-x^2\left(1+\frac{x^2}{2}\right)^{-1}\\
\nonumber&=&2-\ln2-\gamma-\frac{2x^2}{2+x^2}\\
&=&0.7296-x^2+\mathcal{O}(x^4),
\end{eqnarray}
\begin{eqnarray}
\nonumber g(x)&=&\left.\left[\frac{d\ln\tilde{H}_{\mu}^{(1)}(x)}{d\mu}\right]^2\right|_{\mu=3/2}+\left.\frac{d^2\ln\tilde{H}_{\mu}^{(1)}(x)}{d\mu^2}\right|_{\mu=3/2}\\
\nonumber&=&f^2(x)+\left.\frac{d^2\ln\Gamma(x)}{d\mu^2}\right|_{\mu=3/2}-\left.\frac{2}{(1-\mu)^3}\left(\frac{x}{2}\right)^{2}\left[1-\frac{1}{1-\mu}\left(\frac{x}{2}\right)^{2}\right]^{-1}\right|_{\mu=3/2}\\
\nonumber&&-\left.\frac{1}{(1-\mu)^4}\left(\frac{x}{2}\right)^{4}\left[1-\frac{1}{1-\mu}\left(\frac{x}{2}\right)^{2}\right]^{-2}\right|_{\mu=3/2}\\
\nonumber&=&f^2(x)+\psi^{(1)}\left(\frac{3}{2}\right)+4x^2\left(1+\frac{x^2}{2}\right)^{-1}-x^4\left(1+\frac{x^2}{2}\right)^{-2}\\
\nonumber&=&\left(2-\ln2-\gamma-\frac{2x^2}{2+x^2}\right)^2+\frac{\pi^2}{2}-4+\frac{8x^2}{2+x^2}-\frac{4x^4}{(2+x^2)^2}\\
\nonumber&=&\frac{\pi^2}{2}+(\ln2+\gamma)\left(-4+\ln2+\gamma+\frac{4x^2}{2+x^2}\right)\\
&=&1.4672+2.5407x^{2}+\mathcal{O}(x^4).
\end{eqnarray}

\section{Integration of super-horizon evolution equations}\label{app-int}
In subsection \ref{subsect-sup}, we derived the evolution equations \eqref{dQdvs} for super-Hubble perturbations and wrote their solution in the form \eqref{Qvs} with an undetermined function $Y$. Actually, as indicated in references \cite{vandeBruck:2014ata,Avgoustidis:2011em}, equations \eqref{dQdvs} can be integrated completely, yielding the formal solution
\begin{eqnarray}\label{Qvs-dBR}
\nonumber Q_{\sigma}(N)&\simeq&e^{\int^{N}_{N_{\ast}}AdN}\left(Q_{\sigma\ast}+\delta s_{\ast}\int^{N}_{N_{\ast}}Be^{\tilde{\gamma}}dN\right),\\
\delta s(N)&\simeq&\delta s_{\ast}e^{\int^{N}_{N_{\ast}}DdN},
\end{eqnarray}
in which $\tilde{\gamma}=\int^{N}_{N_{\ast}}(D-A)dN$. Correspondingly, the power spectra and correlation spectrum on super-Hubble scales take the form
\begin{eqnarray}
\mathcal{P}_{\mathcal{R}}(N)&\simeq&\mathcal{P}_{\mathcal{R}\ast}+\mathcal{P}_{\mathcal{S}\ast}\left(\int^{N}_{N_{\ast}}Be^{\tilde{\gamma}}dN\right)^2+2\mathrm{Re}(\mathcal{C}_{\mathcal{RS}\ast})\int^{N}_{N_{\ast}}Be^{\tilde{\gamma}}dN,\label{PR-dBR}\\
\mathcal{C}_{\mathcal{RS}}(N)&\simeq&\mathcal{C}_{\mathcal{RS}\ast}e^{\tilde{\gamma}}+\mathcal{P}_{\mathcal{S}\ast}e^{\tilde{\gamma}}\int^{N}_{N_{\ast}}Be^{\tilde{\gamma}}dN,\label{CRS-dBR}\\
\mathcal{P}_{\mathcal{S}}(N)&\simeq&\mathcal{P}_{\mathcal{S}\ast}e^{2\tilde{\gamma}},\label{PS-dBR}
\end{eqnarray}
This form of spectra is inconvenient for analytical calculations but useful for numerical computations.

\begin{figure}
  \centering
  \includegraphics[width=0.33\textwidth]{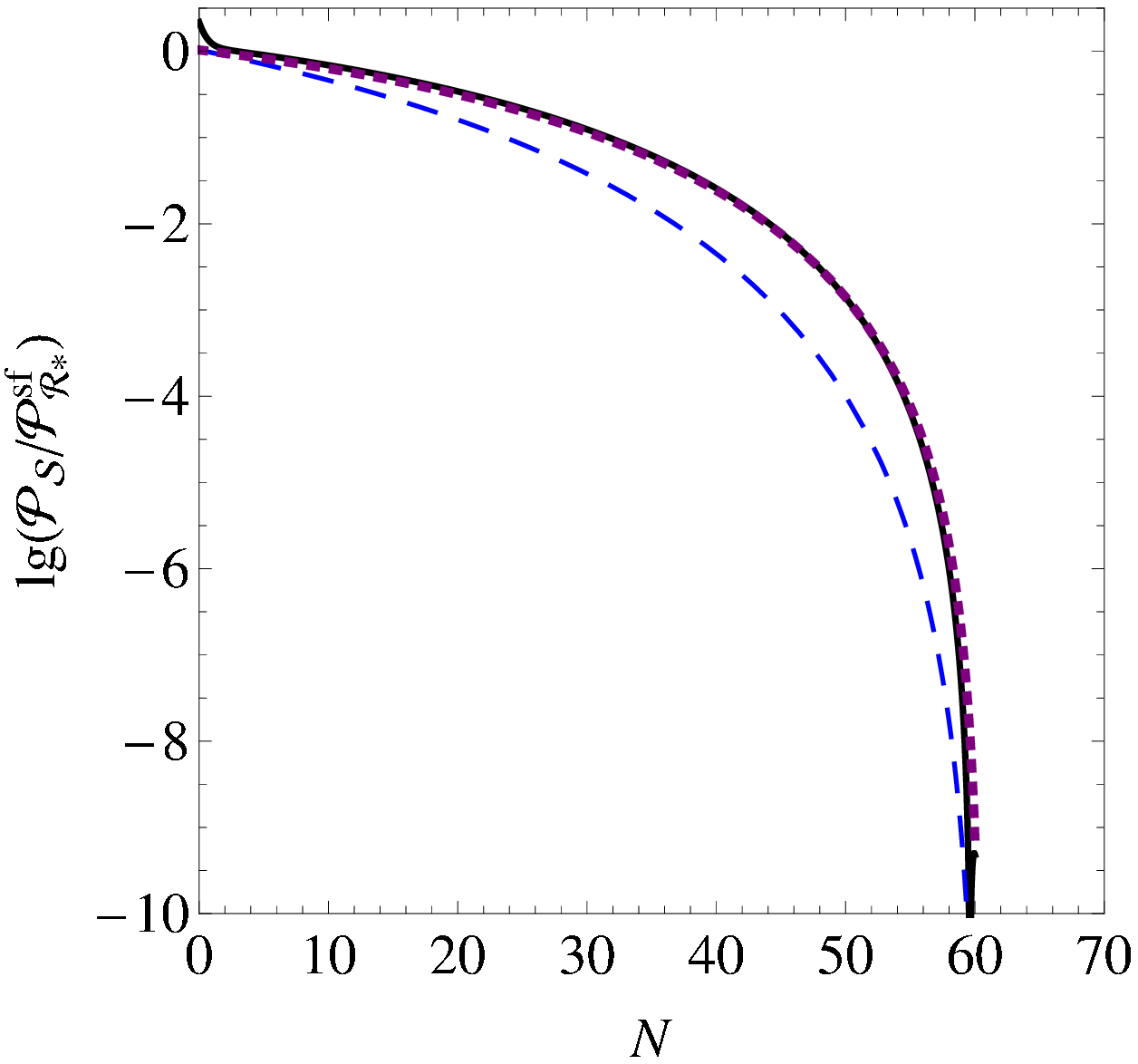}\includegraphics[width=0.33\textwidth]{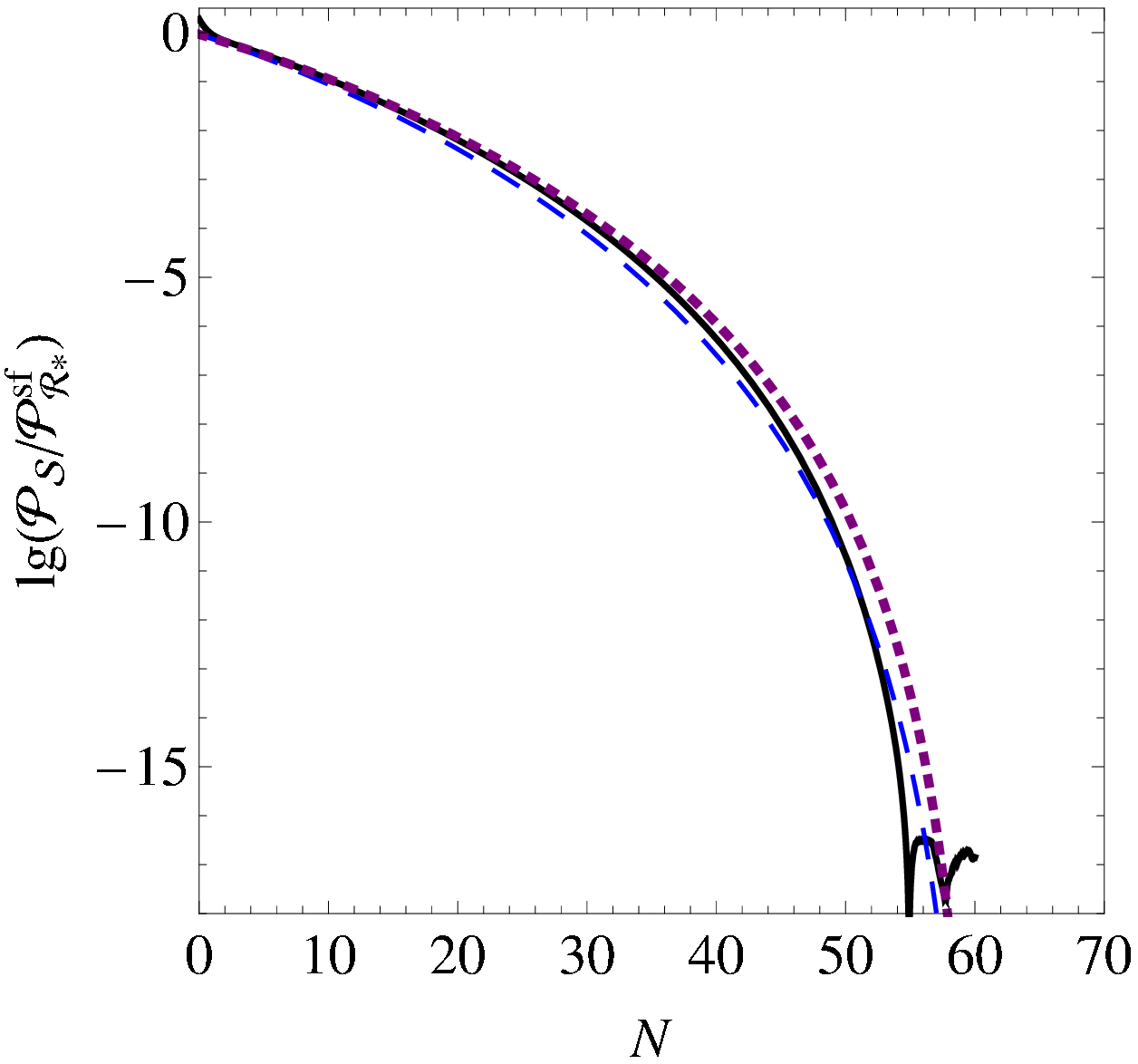}\includegraphics[width=0.33\textwidth]{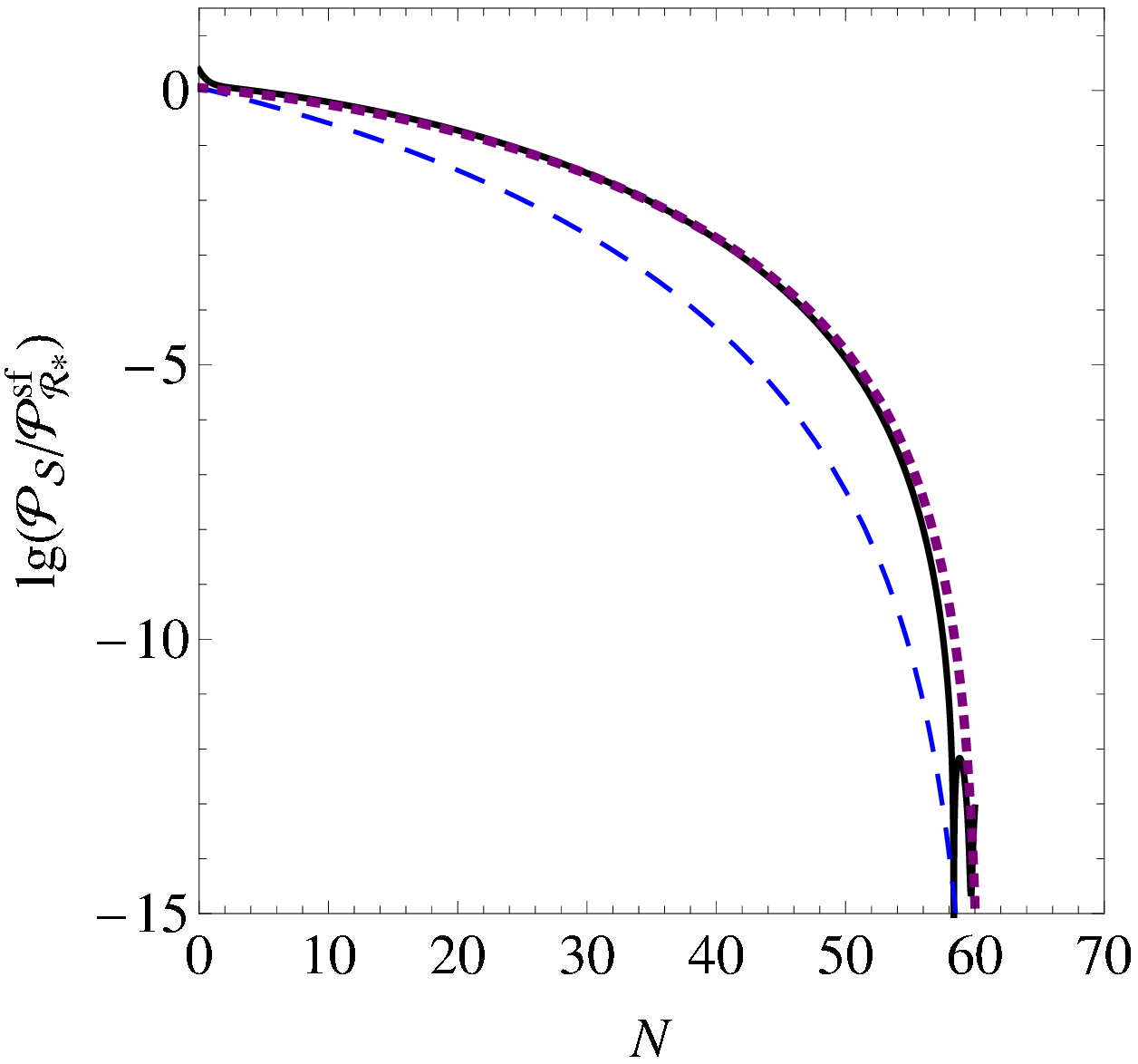}\\
  \caption{(color online). Numerical integration predictions against the numerical result for the power spectrum of entropy perturbation $\mathcal{P_{S}}$ in model \eqref{model} with parameters given by \eqref{param}. Thick solid lines depict the numerical result. Numerical integration predictions \eqref{PS-dBR} are shown by blue dashed lines with $D_{dBR}$ in equation \eqref{D-dBR} and purple dotted lines with $D$ in equation \eqref{ABD}.}\label{fig-PS-dBR}
\end{figure}

As discussed in the main body, the expression of $C_{ss}$ in this paper is different from the one in reference \cite{vandeBruck:2014ata}. Such a difference leads to different values of $D$ given by equations \eqref{ABD}, \eqref{D-dBR}. Substituted into the formulae above, they yield quantitatively different power spectra. In section \ref{sect-num}, we have made a comparison between them by an analytical approximate approach for three benchmark models. In this appendix, we would like to make a more precise comparison by numerical integration. This will help us to confirm the expression of $C_{ss}$. Since there is no difference in $A$ or $B$, it is enough to restrict to the power spectrum of entropy perturbation \eqref{PS-dBR}. We plug equations \eqref{ABD} and \eqref{D-dBR} respectively into formula \eqref{PS-dBR}, and then perform the integral numerically. The results, which we call the numerical integration results, are plotted in figure \ref{fig-PS-dBR} against purely numerical result \cite{Tsujikawa:2002qx} of section \ref{sect-num}. The purely numerical result is shown by thick solid lines as we have done figures \ref{fig-PS}. The numerical integration result with $D$, depicted by purple dotted lines, behaves slightly better than the numerical integration result with $D_{dBR}$ i.e. the blue dashed lines.

\section{Partial analytical results to order $\xi^3$}\label{app-o3}
In this section, we will give some estimate about the contribution of the terms of order $\xi^3$. Since we take $\xi^2$ to be of the same order as $\epsilon$ in this paper, we should pick up all the terms of $\mathcal{O}\left(\xi^{3},\epsilon\xi,\eta\xi\right)$ in this section for consistency. The extension of equations \eqref{ww}, \eqref{dvs2} to higher orders is very cumbersome, so we will report the $\mathcal{O}\left(\xi^{3},\epsilon\xi,\eta\xi\right)$ corrections to equations \eqref{Csr}, \eqref{Mtilde} only, which lead to corrections of the same order to the spectra \eqref{PRstar}, \eqref{PSstar}, \eqref{CRSstar}. The final results are summarized below. As they should be, all of the results are suppressed by a factor $\xi$ in comparison with our results in the main text.

\begin{eqnarray}\label{Csr3}
\nonumber C_{\sigma\sigma}&=&\mathrm{terms~in~equation~\eqref{Csr}}+H^2\xi  s_{\theta}\left[4\epsilon s_{\theta}c_{\theta}+\eta_{\sigma s}(3s_{\theta}^{2}-2)+\frac{2}{3}\xi^2s_{\theta}^{3}c_{\theta}(1-2s_{\theta}^{2})\right]+\mathcal{O}\left(\epsilon^{2},\eta^{2},\xi^{4},\epsilon\eta,\epsilon\xi^{2},\eta\xi^{2}\right),\\
\nonumber C_{\sigma s}&=&\mathrm{terms~in~equation~\eqref{Csr}}+H^2\xi\left[-2\epsilon s_{\theta}^{3}+2\eta_{\sigma s} c_{\theta}(2s_{\theta}^{2}-1)+\frac{4}{3}\xi^2s_{\theta}^{3}c_{\theta}^{2}(1-2s_{\theta}^{2})\right]+\mathcal{O}\left(\epsilon^{2},\eta^{2},\xi^{4},\epsilon\eta,\epsilon\xi^{2},\eta\xi^{2}\right),\\
\nonumber C_{ss}&=&\mathrm{terms~in~equation~\eqref{Csr}}+H^2\xi s_{\theta}\left[\eta_{\sigma s}(1+s_{\theta}^{2})+\frac{2}{3}\xi^2s_{\theta}c_{\theta}(1-3s_{\theta}^{2})\right]+\mathcal{O}\left(\epsilon^{2},\eta^{2},\xi^{4},\epsilon\eta,\epsilon\xi^{2},\eta\xi^{2}\right),\\
C_{s\sigma}&=&\mathrm{terms~in~equation~\eqref{Csr}}+H^2\xi s_{\theta}^{2}\left(-2\epsilon s_{\theta}-2\eta_{\sigma s}c_{\theta}+\frac{8}{3}\xi^2s_{\theta}^{3}c_{\theta}^{2}\right)+\mathcal{O}\left(\epsilon^{2},\eta^{2},\xi^{4},\epsilon\eta,\epsilon\xi^{2},\eta\xi^{2}\right).
\end{eqnarray}
\begin{eqnarray}\label{Mtilde3}
\nonumber\tilde{\mathbf{M}}&=&-\mathbf{E^{\mathrm{2}}+E+M}\\
\nonumber&=&\mathrm{terms~in~equation~\eqref{Mtilde}}+\mathcal{O}\left(\epsilon^{2},\eta^{2},\xi^{4},\epsilon\eta,\epsilon\xi^{2},\eta\xi^{2}\right)+\left(
     \begin{array}{cc}
       \frac{2}{3}\xi^3s_{\theta}^{4} c_{\theta} & \frac{8}{3}\xi^3s_{\theta}^{3}c_{\theta}^{4} \\
       \frac{4}{3}\xi^3s_{\theta}^{3}c_{\theta}^{2}(2s_{\theta}^{2}-1) & \frac{2}{3}\xi^3s_{\theta}^{2}c_{\theta}^{3} \\
     \end{array}
   \right)\\
&&+\left(
     \begin{array}{cc}
       \xi s_{\theta}\left[10\epsilon s_{\theta}c_{\theta}-\eta_{\sigma s}(1+c_{\theta}^{2})\right] & \xi\left[-11\epsilon s_{\theta}^{3}+2\eta_{\sigma\sigma}s_{\theta}^{3}-2\eta_{ss}s_{\theta}c_{\theta}^{2}+2\eta_{\sigma s}c_{\theta}(2s_{\theta}^{2}-1)\right] \\
       \xi s_{\theta}\left[-5\epsilon s_{\theta}^{2}-2\eta_{\sigma\sigma}s_{\theta}^{2}+2\eta_{ss}c_{\theta}^{2}-2\eta_{\sigma s}s_{\theta}c_{\theta}\right] & \xi c_{\theta}\left[-6\epsilon(1+s_{\theta}^{2})+\eta_{\sigma s}s_{\theta}c_{\theta}\right] \\
     \end{array}
   \right),
\end{eqnarray}

\begin{eqnarray}\label{PRstar3}
\nonumber\bar{\mathcal{P}}_{\mathcal{R}}&=&\mathrm{terms~in~equation~\eqref{PRstar}}+\mathcal{O}\left(\lambda_{A}^{3}\right)+\mathcal{O}\left(\epsilon^{2},\eta^{2},\xi^{4},\epsilon\eta,\epsilon\xi^{2},\eta\xi^{2}\right)\\
\nonumber&&+\left(\frac{H^2_{\ast}}{2\pi\dot{\sigma}_{\ast}}\right)^{2}(1+k^{2}\tau^{2})\frac{\xi_{\ast}s_{\theta\ast}}{4s_{\theta\ast}^{6}+\left(2s_{\theta\ast}^{2}+1\right)^2c_{\theta\ast}^{2}}\biggr\{\frac{1}{24}\pi^2\xi_{\ast}^2s_{\theta\ast}^{5}c_{\theta\ast}(2c_{2\theta\ast}+3c_{4\theta\ast}-17)\\
\nonumber&&+\frac{1}{36}(3c_{2\theta\ast}-5)\left[6\eta_{\sigma s\ast}c_{2\theta\ast}-30\eta_{\sigma s\ast}-\xi_{\ast}^2s_{4\theta\ast}+2\left(6\eta_{\sigma\sigma\ast}+\xi_{\ast}^2-6\epsilon_{\ast}\right)s_{2\theta\ast}\right]f(-k\tau)\\
\nonumber&&+\frac{1}{192}\left[\xi_{\ast}^2(-10c_{3\theta\ast}-6c_{5\theta\ast}+3c_{7\theta\ast}+c_{9\theta\ast})+12\left(-56\eta_{\sigma\sigma\ast}+\xi_{\ast}^2+168\epsilon_{\ast}\right)c_{\theta\ast}\right]s_{\theta\ast}f^2(-k\tau)\\
\nonumber&&+\frac{1}{96}\left[\xi_{\ast}^2(-62c_{3\theta\ast}+14c_{5\theta\ast}+5c_{7\theta\ast}-c_{9\theta\ast})+4\left(-84\eta_{\sigma\sigma\ast}+11\xi_{\ast}^2+252\epsilon_{\ast}\right)c_{\theta\ast}\right]s_{\theta\ast}g(-k\tau)\\
&&+\frac{1}{2}[\eta_{\sigma s\ast}(13s_{\theta\ast}-3s_{3\theta\ast})+3(\eta_{\sigma\sigma\ast}-3\epsilon_{\ast})c_{3\theta\ast}]s_{\theta\ast}\left[f^2(-k\tau)+g(-k\tau)\right]\biggr\},
\end{eqnarray}
\begin{eqnarray}\label{PSstar3}
\nonumber\bar{\mathcal{P}}_{\mathcal{S}}&=&\mathrm{terms~in~equation~\eqref{PSstar}}+\mathcal{O}\left(\lambda_{A}^{3}\right)+\mathcal{O}\left(\epsilon^{2},\eta^{2},\xi^{4},\epsilon\eta,\epsilon\xi^{2},\eta\xi^{2}\right)\\
\nonumber&&+\left(\frac{H^2_{\ast}}{2\pi\dot{\sigma}_{\ast}}\right)^{2}(1+k^{2}\tau^{2})\xi_{\ast}\biggr\{\frac{1}{12}\pi^2\xi_{\ast}^2s_{\theta\ast}^{6}(5c_{\theta\ast}+c_{3\theta\ast})\\
\nonumber&&-\frac{2}{9}s_{\theta\ast}c_{\theta\ast}^{2}\left(3\eta_{\sigma s\ast}+\xi_{\ast}^2s_{2\theta\ast}\right)f(-k\tau)+\frac{1}{3}xi^2s_{\theta\ast}^{6}(5c_{\theta\ast}+c_{3\theta\ast})f^2(-k\tau)\\
&&+\left[\left(5\epsilon_{\ast}-5\eta_{ss\ast}+\frac{4}{3}\xi_{\ast}^2\right)c_{\theta\ast}-4\eta_{\sigma s\ast}s_{\theta\ast}^{3}+(\eta_{ss\ast}-\epsilon_{\ast})c_{3\theta\ast}\right]\left[\frac{1}{2}g(-k\tau)+\frac{1}{2}f^2(-k\tau)-\frac{1}{3}f(-k\tau)\right]\biggr\},
\end{eqnarray}
\begin{eqnarray}\label{CRSstar3}
\nonumber\bar{\mathcal{C}}_{\mathcal{RS}}&=&\mathrm{terms~in~equation~\eqref{CRSstar}}+\mathcal{O}\left(\lambda_{A}^{3}\right)+\mathcal{O}\left(\epsilon^{2},\eta^{2},\xi^{4},\epsilon\eta,\epsilon\xi^{2},\eta\xi^{2}\right)\\
\nonumber&&+\left(\frac{H^2_{\ast}}{2\pi\dot{\sigma}_{\ast}}\right)^{2}(1+k^{2}\tau^{2})\xi_{\ast}\biggr\{\frac{1}{24}\pi^2\xi_{\ast}^2s_{\theta\ast}^{3}c_{\theta\ast}^{2}(c_{4\theta\ast}-7)\\
\nonumber&&+\frac{1}{18}\left[3\eta_{\sigma s\ast}(7c_{\theta\ast}+c_{3\theta\ast})+12(\eta_{ss\ast}+\eta_{\sigma\sigma\ast}-2\epsilon_{\ast})s_{\theta\ast}^{3}-8\xi_{\ast}^2s_{\theta\ast}^{3}c_{\theta\ast}^{2}\right]f(-k\tau)\\
\nonumber&&+\frac{1}{24}\xi_{\ast}^2s_{\theta\ast}^{3}(-13c_{2\theta\ast}+2c_{4\theta\ast}+c_{6\theta\ast}-14)f^2(-k\tau)\\
\nonumber&&+\left[(-\eta_{ss\ast}-\eta_{\sigma\sigma\ast}+4\epsilon_{\ast})s_{\theta\ast}^{3}-\eta_{\sigma s\ast}c_{\theta\ast}\right]\left[f^2(-k\tau)+g(-k\tau)\right]\\
\nonumber&&-\frac{1}{12}i\pi\left[\eta_{\sigma s\ast}(c_{\theta\ast}+3c_{3\theta\ast})+8\eta_{ss\ast}s_{\theta\ast}c_{\theta\ast}^{2}-\xi_{\ast}^2s_{\theta\ast}^{2}c_{\theta\ast}s_{4\theta\ast}+4(3\epsilon_{\ast}-2\eta_{\sigma\sigma\ast})s_{\theta\ast}^{3}\right]\\
&&+\frac{1}{3}i\pi\xi_{\ast}^2s_{\theta\ast}^{3}c_{\theta\ast}^{2}(c_{2\theta\ast}+2)f(-k\tau)\biggr\}.
\end{eqnarray}

We should emphasize that the exposed terms in equations \eqref{PRstar3}, \eqref{PSstar3}, \eqref{CRSstar3} are incomplete, because we did not worked out the $\mathcal{O}\left(\lambda_{A}^{3}\right)$ terms, that is, the higher order corrections to equation \eqref{ww}. These corrections are also of order $\xi^3$ in view of equations \eqref{eigen}. We can expect that the full corrections are still suppressed by a factor $\xi$ in comparison with our results in the main text.

\end{document}